\documentclass[preprint,aps,amsmath,pre]{revtex4}

\usepackage{natbib}

\usepackage{graphicx}
\usepackage{float}
\usepackage{amssymb}

\begin{document}

\title{Complex dynamics in coevolution models with
ratio-dependent functional response
}
\author{Per Arne Rikvold
}

\affiliation{
Center for Materials Research and Technology,
National High Magnetic Field Laboratory, and Department of Physics,\\
Florida State University, Tallahassee, Florida 32306-4350, USA\\
Tel.: +1-850-644-6814, Fax.: +1-850-644-6504, E-mail: prikvold@fsu.edu
}

\begin{abstract}
We explore the complex dynamical behavior of two simple predator-prey 
models of biological coevolution that on the ecological level account 
for interspecific and intraspecific competition, as well
as adaptive foraging behavior. The underlying individual-based population 
dynamics are based on a ratio-dependent functional response 
[W.M. Getz, J.\ theor.\ Biol. \textbf{108}, 623 (1984)]. 
Analytical results for fixed-point population sizes in some simple 
communities are derived and discussed. 
In long kinetic Monte Carlo simulations we 
find quite robust, approximate $1/f$ noise in species diversity and
population sizes, as well as power-law distributions for the
lifetimes of individual species and the durations of
periods of relative evolutionary stasis. 
Adaptive foraging enhances coexistence of species and
produces a metastable low-diversity
phase and a stable high-diversity phase. 

\end{abstract}

\maketitle

\section{Introduction}
\label{sec:int}

In recent years, there has been a growing  
recognition that processes at the ecological and evolutionary scales can
be strongly linked \citep{DROS01B,THOM98,THOM99,YOSH03}.
As a consequence, several approaches have been proposed, which 
model the complex process of coevolution in a fitness
landscape that changes with the composition of the community,
while spanning wide ranges of both temporal and taxonomic scales.
Early steps in this direction were simulations of parapatric and
sympatric speciation \citep{CROS70} and the coupled $NK$ model with
population dynamics \citep{KAUF93,KAUF91}.
More recent contributions include the webworld model
\citep{CALD98,DROS01B,DROS04,QUIN05B,QUIN05}, 
the simple origination/extinction model of \citet{NUNE99}, 
the speciation model of \citet{ROSS05}, 
the matching model of \citet{ROSS06}, the family of models 
introduced by \citet{YOSH08},
the individual-based tangled-nature model
\citep{CHRI02,COLL03,HALL02}
and simplified versions of the latter
\citep{RIKV05A,RIKV06,RIKV06B,RIKV03,SEVI06,ZIA04}, 
as well as network models
\citep{CHOW05,CHOW03A,QIN07}.
Recently, large individual-based simulations have also been performed of
parapatric and sympatric speciation \citep{GAVR98,GAVR00} and of
adaptive radiation \citep{GAVR05}.

Many of the models mentioned above are deliberately simple, aiming to
elucidate {\it universal\/} features that are largely independent of the
finer details of the ecological interactions and the evolutionary
mechanisms. While valuable from this point of view, the departure of
many such models from mechanisms usually included in models
of population dynamics and evolution has limited their acceptance in the
biological community. A case in point are the  
tangled-nature model \citep{CHRI02,COLL03,HALL02}
and its simplifications 
\citep{RIKV05A,RIKV06,RIKV06B,RIKV03,SEVI06,ZIA04}. 
Here, we therefore introduce a modification of the latter, in which 
intra- and interspecific competition and adaptive foraging are introduced
through ratio-dependent functional response functions. 

Our motivation for this study
is primarily a desire to understand the extent to which the 
long-time dynamics of complex coevolution models depend on details of the 
population dynamics. By using individual-based models with mutations, 
we avoid introducing 
the artificial separation between speciations and population dynamics
inherent in all species-level models, including those mentioned above 
\citep{CALD98,DROS01B,DROS04,QUIN05B,QUIN05,NUNE99,ROSS05,ROSS06,YOSH08}. 
Despite being individual-based, our models enable fast simulation of large 
communities over time scales of tens of millions of generations. An important 
question that one would like to answer in the future is whether the 
avalanche-like mass extinctions observed 
in the fossil record are due to intrinsic fluctuations of the 
nonlinear dynamics [so-called self-organized criticality or SOC 
\citep{NEWM96}], or to external perturbations such as asteroid impacts, 
volcanic eruptions, or climate changes, or to a combination 
of intrinsic and extrinsic causes \citep{NEWM03}. In order to successfully 
address this question, it is necessary to understand better the influence 
that the population dynamics have on the intrinsic fluctuations in well-defined 
model systems. We find that the dynamics of the functional-response 
models studied in this paper differ from the tangled-nature type models 
studied earlier and also 
from each other, depending on whether or not adaptive foraging is 
implemented. 

The remainder of this paper is organized as follows. 
The model without adaptive foraging is defined in Sec.~\ref{sec:mod}. 
For this model, analytical results for simple communities  
are derived and discussed in Sec.~\ref{sec:Ana}, 
and kinetic Monte Carlo simulations of multispecies communities with 
mutations are performed and analyzed in Sec.~\ref{sec:Sim1}. 
Adaptive foraging is introduced and investigated in Sec.~\ref{sec:adap}, 
both by numerical solution of the steady-state equations
for two-species communities, and by long-time
kinetic Monte Carlo simulations for evolving multispecies communities. 
A summary and conclusions are given in Sec.~\ref{sec:conc}. 

\section{Models}
\label{sec:mod}

We recently performed detailed analytical and simulational studies 
of the long-time dynamics and community structures of simplified
tangled-nature models 
\citep{RIKV05A,RIKV06,RIKV06B,RIKV03,SEVI06,ZIA04}.
In particular, the behaviors of mutualistic and predator-prey versions
were compared by \citet{RIKV06}, and community structures of the
latter were compared with data from real ecosystems by \citet{RIKV06B}. 
Here we first describe features of these models that are shared by the new
models that will be introduced below. More detailed descriptions 
of our previous models are given by \citet{RIKV06} and \citet{RIKV06B}. 

\subsection{Shared features}
\label{sec:shared}

The mechanism for selection between several interacting species is
provided by   
the reproduction rates in an individual-based population 
dynamics with nonoverlapping generations. 
At the end of each generation, each individual of species $I$ reproduces
asexually, giving birth to a fixed number 
$F$ of offspring with probability $P_I$ before dying, or it dies
without offspring with probability $(1-P_I)$. Each $P_I$ 
depends on the set $\{n_J(t)\}$ of
population sizes of all the species resident in the community in
generation $t$ through 
interspecies interactions and other model parameters as
described below. The interactions are determined by the random {\it
interaction matrix\/} $\bf M$ \citep{SOLE96}, which is constructed at
the beginning of a simulation run, and thereafter kept constant.
If $M_{IJ}$ is positive and $M_{JI}$
negative, then $I$ is a predator and $J$ its prey, and vice versa. If
both matrix elements are positive, the relationship is a 
mutualistic one, while both negative indicate an antagonistic
relationship. 

The species are defined by a haploid, binary ``genome" of
length $L$, as in Eigen's model for molecular 
evolution \citep{EIGE71,EIGE88}, and 
the potential species are identified by the index $I \in [0,2^L-1]$. 
Typically, only $\mathcal{N}(t) \ll 2^L$
of these potential species are actually 
resident in the community at any one time $t$.

New species enter the community as each offspring organism may mutate   
with a small probability $\mu$. Mutation consists in flipping a
randomly chosen bit in the genome, and a mutated individual
is assumed to belong to a different species than its parent, with
different properties. Genotype and phenotype are thus in one-to-one
correspondence in these models.
This is a highly idealized picture, which is introduced
to maximize the pool of different species available within the
computational resources. The approximation is justified by
a computational study, in which species differing by as many as
$L/2$ bits have correlated properties \citep{SEVI06}.
Remarkably, it was shown that even strong correlations between the
phenotypes of parents and offspring are relatively
unimportant for the long-time dynamical properties. 

Regardless of the functional form of $P_I$, 
the time development of the mean population sizes, 
$\langle n_I(t) \rangle$, is described by a set of coupled difference equations,
\begin{eqnarray}
\langle n_I(t+1) \rangle
&=& \langle n_I(t) \rangle FP_I(\{\langle n_J(t)\rangle \})[1-\mu ]
\nonumber\\
&& +(\mu/L)F\sum_{K(I)}\langle n_{K(I)}(t)
\rangle P_{K(I)}(\{\langle n_J(t) \rangle \}) 
\;,
\label{eq:MF}
\end{eqnarray}
where $K(I)$ represents the $L$ species that can be generated from $I$
by a single mutation.

\subsection{Simplified tangled-nature models}
\label{sec:tana}

In these models,
the reproduction probability for an individual of
species $I$ is given by the nonlinear function 
\begin{equation}
P_I(t) = \frac{1}{1 + \exp[-\Delta_I(R,\{n_J(t)\})]} \;,
\label{eq:PI}
\end{equation}
where $R$ is an external resource that is renewed at the 
same level each generation. The function $\Delta_I$ is given by  
\begin{equation}
\Delta_I(R,\{n_J(t)\}) = - b_I + {\eta_I R \over N_{\rm tot}(t)} 
+ \sum_J {M_{IJ} n_J(t) \over N_{\rm tot}(t)} - {N_{\rm tot}(t) \over N_0}
\;.
\label{eq:Delta}
\end{equation}
Here $b_I$ is a ``reproduction cost" (always
positive), and $\eta_I$ (positive for primary producers or autotrophs,
and zero for consumers or heterotrophs) is the ability of
individuals of species $I$ to utilize the external resource $R$,
while $N_0$ is an environmental carrying capacity \citep{MURR89} 
[a.k.a.\ Verhulst factor \citep{VERH1838}].
The total population size is $N_{\rm tot}(t) = \sum_J n_J(t)$.

Two versions of this model were studied in earlier work. In the
first version
there is no external resource or birth cost, and the off-diagonal
elements of $\bf M$ are stochastically independent and uniformly
distributed over $[-1,+1]$, while the diagonal elements are zero. 
This model evolves toward mutualistic communities \citep{BASC06}, 
in which all species are connected by asymmetric, mutually 
positive interactions \citep{RIKV03,RIKV06,SEVI06,ZIA04}. 

In the predator-prey version of the model, a small minority
of the potential species (typically 5\%) are primary producers, while the
rest are consumers. The off-diagonal part of the interaction matrix is
antisymmetric, with the additional restriction that a producer
cannot also prey on a consumer \citep{RIKV05A,RIKV06,RIKV06B}. 
In simulations we took $b_I$ and the nonzero $\eta_I$ as
independent and uniformly distributed on $(0,+1]$. 
This model generates simple food webs with up to three 
trophic levels \citep{RIKV05A,RIKV06,RIKV06B}.  

Both of these models provide interesting results, which include
intermittent dynamics with power spectral densities (PSDs) of diversities
and population sizes that exhibit approximate $1/f$ noise, 
as well as power-law distributions for the lifetimes of individual
species and the duration of quiet periods of relative evolutionary stasis.
From a theoretical point of view they
also have the great advantage that the mean-field equation for the
steady-state average population sizes, Eq.~(\ref{eq:MF}), 
in the absence of mutations reduces to a
set of quadratic equations (linear if $N_0 = \infty$) and thus
can easily be solved exactly \citep{RIKV06,RIKV06B,RIKV03,ZIA04}. 
The models thus provide useful benchmarks for more realistic, 
but generally highly nonlinear models, like the ones defined below. 

The population dynamics
defined by Eqs.~(\ref{eq:PI}) and~(\ref{eq:Delta}) have some less 
realistic features. In particular, by summing over positive and negative
terms in $\Delta_I$, the models enable species with little food to
remain near a steady state if they are also not very popular as
prey, or have very low birth cost. 
Another problem is the {\it ad-hoc\/} nature of the
normalization by the total population size $N_{\rm tot}(t)$ in the
resource and interaction terms in $\Delta_I$. While this is the source
of the models' analytic solvability, it implies an 
indiscriminate, universal competition without regard to whether or
not two species directly utilize the same resources or share a
common predator. The purpose of the present paper is to develop models
with more realistic population dynamics and explore 
their complex dynamics on time scales from ecological (short) 
to evolutionary (long). 

\subsection{Functional-response model without adaptive foraging}
\label{sec:holl}

Here we develop a model with population
dynamics that include competition between different
predators that prey on the same species, as well as 
intraspecific competition and a saturation
effect expected to occur for a predator with abundant prey. 
In doing so, we retain from the models discussed above the
important role of the interaction matrix $\bf M$, as well as the
mutation process of the 
binary ``genome" and the restriction to nonoverlapping generations.

We first deal with the competition between predator species
by defining the number of
individuals of $J$ that are available as prey for $I$, corrected for
competition from other predator species, as 
\begin{equation}
\hat{n}_{IJ} = \frac{n_I M_{IJ}}{\sum_L^{{\rm pred}(J)} n_L M_{LJ}} n_J \;,
\label{eq:neff}
\end{equation}
where $\sum_L^{{\rm pred}(J)}$ runs over all $L$ 
such that $M_{LJ} > 0$, i.e.,
over all predators of $J$. Thus, 
$\sum_I^{{\rm pred}(J)} \hat{n}_{IJ} = n_J$, and if $I$ is the only
predator consuming $J$, then $\hat{n}_{IJ} = n_J$. 

Analogously, we define the competition-adjusted
external resources available to a producer species $I$ as 
\begin{equation}
\hat{R}_I = \frac{n_I \eta_I}{\sum_L n_L \eta_L} R \;.
\label{eq:reff}
\end{equation}
As in the case of predators, $\sum_I \hat{R}_I = R$, and a
sole producer species has all of the external resources available to
it: $\hat{R}_I = R$. With these definitions, the total,
competition-adjusted resources available for the sustenance of
species $I$ are 
\begin{equation}
\hat{S}_I = \eta_I \hat{R}_I + \sum_J^{{\rm prey}(I)} M_{IJ} \hat{n}_{IJ} 
\;,
\label{eq:SI}
\end{equation}
where $\sum_J^{{\rm prey}(I)}$ runs over all $J$ such that $M_{IJ} >
0$, i.e., over all prey of $I$, and $\eta_I = 0$ if $I$ is a heterotroph. 

A central concept of the model is the {\it functional response\/} of
species $I$ with respect to $J$, $\Phi_{IJ}$ \citep{DROS01B,KREB01}. 
This is the rate at which an individual of species $I$ consumes
individuals of $J$. The simplest functional response corresponds to
the Lotka-Volterra model \citep{MURR89}: $\Phi_{IJ} = n_J$ if $M_{IJ} > 0$ 
and 0 otherwise. However, it is reasonable to expect that the
consumption rate should saturate in the presence of very abundant
prey \citep{KREB01}. For ecosystems consisting of a single pair of
predator and prey, or a simple chain reaching from a bottom-level
producer through intermediate species to a top predator, the most
common forms of functional response are due to Holling \citep{KREB01}.
For more complicated, interconnected food webs, a number of
functional forms have been proposed in the recent 
literature \citep{DROS01B,DROS04,KUAN02,MART06,QUIN05B,QUIN05,SKAL01}, 
but there is as yet no agreement about a standard form.
Here we choose the ratio-dependent 
\citep{DROS01B,DROS04,FILO08,QUIN05B,QUIN05,RESI95} 
Holling Type II form \citep{KREB01}, originally introduced by 
\citet{GETZ84}, 
\begin{equation}
\Phi_{IJ} = \frac{M_{IJ} \hat{n}_{IJ}}{\lambda \hat{S}_I + n_I} \;,
\label{eq:PhiIJ}
\end{equation}
where $\lambda \in (0,1]$ is the metabolic efficiency of converting
prey biomass to predator offspring. 
The ratio dependence corresponds to intraspecific competition \citep{GETZ84}. 
Analogously, the functional response of a producer species 
$I$ toward the external resource $R$ is 
\begin{equation}
\Phi_{IR} = \frac{\eta_I \hat{R}_{I}}{\lambda \hat{S}_I + n_I} \;.
\label{eq:PhiIR}
\end{equation}
In both cases, if $\lambda \hat{S}_I \ll n_I$, then the consumption
rate equals the resource ($M_{IJ} \hat{n}_{IJ}$ or 
$\eta_I \hat{R}_{I}$) divided by the number of individuals of $I$,
thus expressing intraspecific competition for scarce
resources. In the opposite limit, $\lambda \hat{S}_I \gg n_I$, 
the consumption rate is proportional to the ratio of the specific,
competition-adjusted resource to the competition-adjusted total
available sustenance, $\hat{S}_I$. The total consumption rate for
an individual of $I$ is therefore 
\begin{equation}
C_I = \Phi_{IR} + \sum_J^{{\rm prey}(I)} \Phi_{IJ} 
= \frac{\hat{S}_I}{\lambda
\hat{S}_I + n_I}
=
\left\{
\begin{array}{lll}
{ \hat{S}_I}/{n_I} & \mbox{for} & \lambda \hat{S}_I \ll n_I \nonumber\\
{1}/{\lambda}     & \mbox{for} & \lambda \hat{S}_I \gg n_I 
\end{array}
\right.
\;.
\label{eq:CI}
\end{equation}
The birth probability 
is assumed to be proportional to the consumption rate, 
\begin{equation}
B_I = \lambda C_I \in [0,+1] \;,
\label{eq:BI}
\end{equation}
while the probability that an individual of $I$ 
avoids death by predation until attempting to reproduce at the end of 
the generation is 
\begin{equation}
A_I = 1 - \sum_J^{{\rm pred}(I)} \Phi_{JI} \frac{n_J}{n_I} \;.
\label{eq:AI}
\end{equation}
The total reproduction probability for an individual of species $I$
in this model is thus 
\begin{equation}
P_I(t) = A_I(t) B_I(t) \;. 
\label{eq:PI2}
\end{equation}

\section{Analytical Results}
\label{sec:Ana}

The functional-response model defined in Sec.~\ref{sec:holl} 
is much less amenable to analytic treatment than the
models we have considered previously. In particular, the simultaneous 
set of equations,
\begin{equation}
F P_I(\{n_J^*\}) = 1
\label{eq:fix}
\end{equation}
where $F$ is the fecundity,
which defines the fixed-point
solution $\{n_I^*\}$ of Eq.~(\ref{eq:MF}) for multispecies 
communities in the mutation-free limit, 
cannot be solved analytically in general. 
However, some special cases can be solved explicitly. 
Although these analytical solutions are highly model-specific, 
they provide useful insight into some of the simplest effects of 
interspecies interactions and intra- and interspecific competition. 

\subsection{Two competing producers with intraguild predation}
\label{sec:compet}

Consider two producer species characterized by their coupling constants, 
$\eta_1$ and $\eta_2$. In the noninteracting case, $M_{21}=M_{12}=0$, the 
species are subject to competitive exclusion, so that only one species, the 
one with the maximum value of $\eta_I$, can survive with a nonzero fixed-point 
population, $n_{\rm max}^* = \lambda \eta_{\rm max} (F-1) R$. 
The only exception is 
the degenerate case of $\eta_1 = \eta_2$, in which the two species are 
dynamically indistinguishable. 
The property of competitive exclusion in the noninteracting 
limit is shared with the simplified tangled-nature models discussed 
previously, and it carries over to sets 
of any number of noninteracting producers \citep{RIKV06}. 

\begin{figure}[t]
\begin{center}
\vspace*{0.3truecm}
\includegraphics[angle=0,width=.47\textwidth]{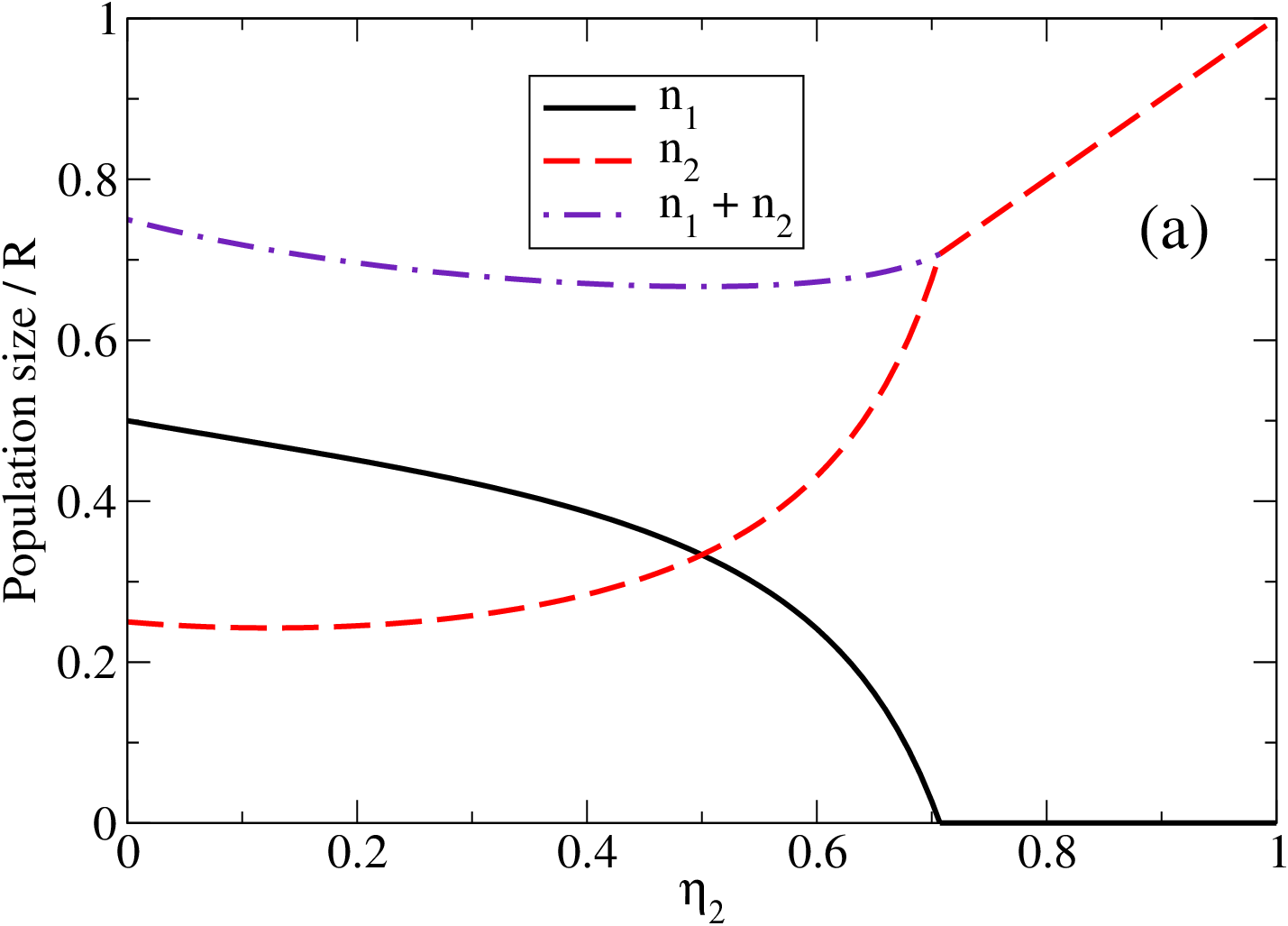}
\hspace{0.5truecm}
\includegraphics[angle=0,width=.47\textwidth]{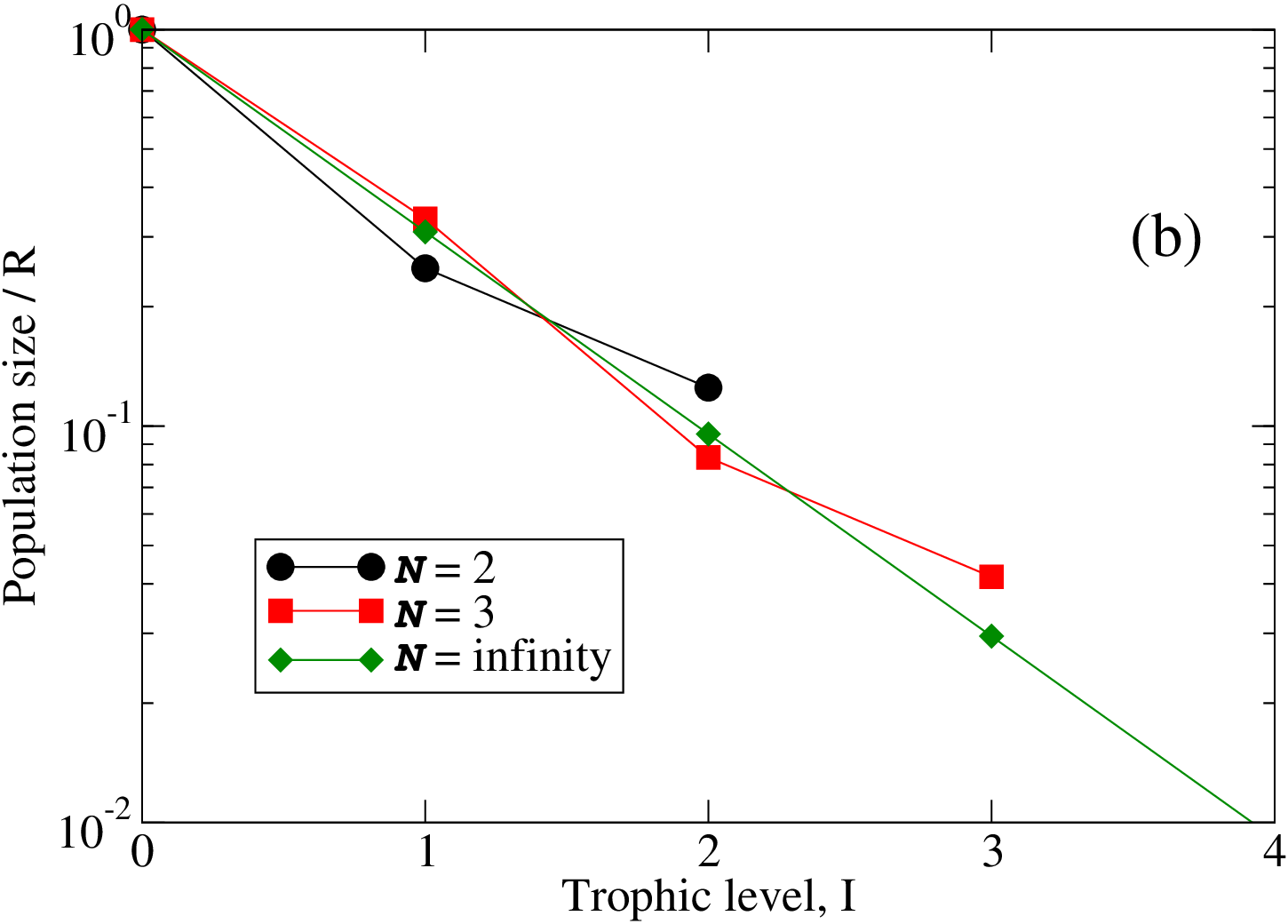}
\end{center}
\caption[]{
{\bf (a)}
Population sizes for two competing producer species with intraguild predation 
in the functional-response model without adaptive foraging,
shown as functions of $\eta_2$. 
The other parameters are $\eta_1 = 1.0$, $M_{21}=0.5$, and $\lambda=1$.
See the text for details. 
{\bf (b)}
Population sizes for food chains of $\mathcal{N}$ species with $\lambda = 1$ 
and $M = 0.5$. 
See the text for details. 
}
\label{fig:levels}
\end{figure}

For the interacting case, $M_{21} \in (0,1]$ (and $M_{12} = - M_{21}$),
which is a simple example of intraguild predation,  
the solution to Eq.~(\ref{eq:fix}) is 
\begin{equation}
n_1^* 
= \frac{\eta_1 \lambda (F-1) [\eta_1^2 (1-M_{21}) - \eta_2^2] (1-M_{21})}
{\eta_1 (1-M_{21}) [\eta_1 + \eta_2 \lambda (F-1) M_{21}] - \eta_2^2} R
\label{eq:n1*}
\end{equation}
and
\begin{equation}
n_2^* 
= \frac{\eta_1^3 \lambda^2 (F-1)^2 (1-M_{21})^2 M_{21}}
{\eta_1 (1-M_{21}) [\eta_1 + \eta_2 \lambda (F-1) M_{21}] - \eta_2^2} R
\;,
\label{eq:n2*}
\end{equation}
as long as both populations are nonnegative. This requires  
$0 \le \eta_2 \le \eta_1 \sqrt{1-M_{21}}$. 
These rather complicated analytical solutions are best interpreted 
graphically as in Fig.~\ref{fig:levels}(a), which shows the case
$\lambda=1$, $\eta_1=1$, and $M_{21}=0.5$. 
Other parameter values give similar results. 
As $\eta_2$ is increased from zero, 
the population of species 2, $n_2$, first decreases weakly as it competes 
directly for resources with species 1, which is its only source of support 
at $\eta_2=0$. Differentiation of the denominator in Eq.~(\ref{eq:n2*}) 
shows that $n_2$ reaches its minimum at 
$\eta_2 = \eta_1 \lambda (1-M)M/2$. For larger $\eta_2$ 
it increases nonlinearly due to the term quadratic in $\eta_2$ in the 
denominator. The combined competition and predation from species 2 causes 
$n_1$ to decrease monotonically, first linearly in $\eta_2$ 
and later nonlinearly until it reaches zero 
at $\eta_2 = \eta_1 \sqrt{1-M_{21}}$. 
For larger $\eta_2$, species 2 completely excludes species 1, and 
the stationary solution is $n_2^* = \lambda \eta_2 (F-1)R$ and $n_1^* = 0$, 
even though $\eta_2$ may still be less than $\eta_1$. 
The two solutions for $n_2$ 
join continuously at $\eta_2 = \eta_1 \sqrt{1-M_{21}}$, 
and there are no other attractive fixed points for the mutation-free dynamics. 
Looking at the total population size, $n_1+n_2$, we find that 
it is a continuous, convex function of $\eta_2$, with a shallow minimum 
at $\eta_2=\eta_1\left\{ 1 - \sqrt{1-\lambda(F-1)(1-M)} \right\}$. 
(The solution $n_1^*=\lambda \eta_1 (F-1)R$ and 
$n_2^* = 0$ is a fixed point as well, but 
it is repulsive under perturbations to $n_2^*$.)

\subsection{$\mathcal{N}$-species food chain}
\label{sec:chain}

The other case, for which the fixed-point population sizes can be found 
relatively easily, is a ``food chain" in which species $I+1$ feeds 
exclusively on the preceding one, $I$. The fixed-point equation 
(\ref{eq:fix}) then takes the form 
\begin{equation}
F \left( 1 - \frac{M_{I+1} n_{I+1}^* }
                  {\lambda M_{I+1} n_{I}^* + n_{I+1}^*}\right)
\frac{\lambda M_I n_{I-1}^*}{\lambda M_I n_{I-1}^* + n_I^*}
= 1
\;,
\label{eq:fpchain}
\end{equation}
where for simplicity we write $M_I$ for $M_{I,I-1}$.  

With boundary conditions $n_0^* = R$ and $n_\infty^* = 0$, 
and with all $M_I =M$ (we define $M_1 = \eta_1$), this set has a 
geometrically decreasing solution of 
the form $n_I^* = R \alpha^I$ with 
\begin{equation}
\alpha = 
\frac{\lambda M}{2} \left\{ \sqrt{F [4M + F(1-M)^2]} + F(1-M) - 2 \right\}
< 1\;.
\label{eq:alpha}
\end{equation}
This solution is included in Fig.~\ref{fig:levels}(b) as the one 
corresponding to $\mathcal{N} = \infty$. [Parameter values for which $\alpha 
\ge 1$ (essentially very large fecundity $F$ combined with $M$ and $\lambda$ 
near unity) are unrealistic.]

If it is instead known that there are $\mathcal{N}$ trophic levels, 
so that $n_{\mathcal{N}+1}^* =0$, then the fixed-point equations can be solved 
analytically in an iterative fashion as follows. 
\begin{enumerate}
\item
Solve the linear fixed-point equation (\ref{eq:fix}) expressing 
$n_\mathcal{N}^* = \mathrm{const.}$,
$$
F \frac{\lambda M_{\mathcal{N}} n_{\mathcal{N}-1}^*}
{\lambda M_{\mathcal{N}} n_{\mathcal{N}-1}^* + n_{\mathcal{N}}^*}
=1 \;,
$$
for $n_{\mathcal{N}-1}^*$. 
\item
Insert the solution for $n_{\mathcal{N}-1}^*$ in terms of 
$n_{\mathcal{N}}^*$ into the next equation in the hierarchy 
(the one expressing $n_{\mathcal{N}-1}^* = \mathrm{const.}$),
$$
F \left( 1 - \frac{M_{\mathcal{N}}n_{\mathcal{N}}^*}
{\lambda M_{\mathcal{N}} n_{\mathcal{N}-1}^* + n_{\mathcal{N}}^*}\right) 
\frac{\lambda M_{\mathcal{N}-1} n_{\mathcal{N}-2}^*}
{\lambda M_{\mathcal{N}-1} n_{\mathcal{N}-2}^* + n_{\mathcal{N}-1}^*}
=1 
\;,
$$
and cancel common factors to get a linear equation for 
$n_{\mathcal{N}-2}^*$ in terms of $n_{\mathcal{N}}^*$. 
\item
Continue until obtaining $n_0^*$ in terms of $n_{\mathcal{N}}^*$.
\item
Rescale the solutions to give $n_0^* = R$. 
\end{enumerate}
With large $\mathcal{N}$ and $I$-independent $M_I$, 
this solution converges toward the decreasing geometric one presented above for 
$I \ll \mathcal{N}$, as shown in Fig.~\ref{fig:levels}(b). 

\begin{figure}[t]
\begin{center}
\vspace*{0.3truecm}
\includegraphics[angle=0,width=.47\textwidth]{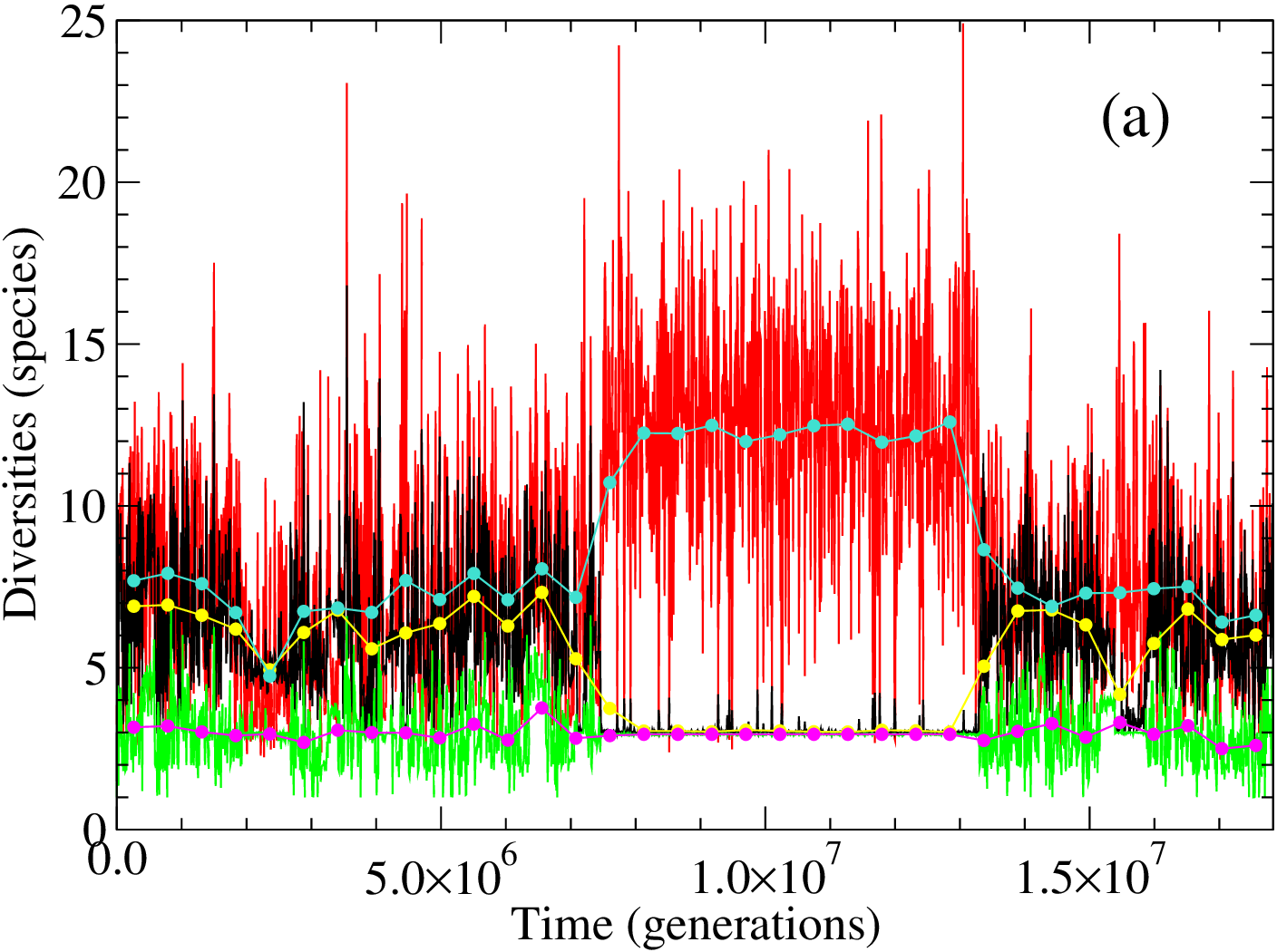}
\hspace{0.5truecm}
\includegraphics[angle=0,width=.47\textwidth]{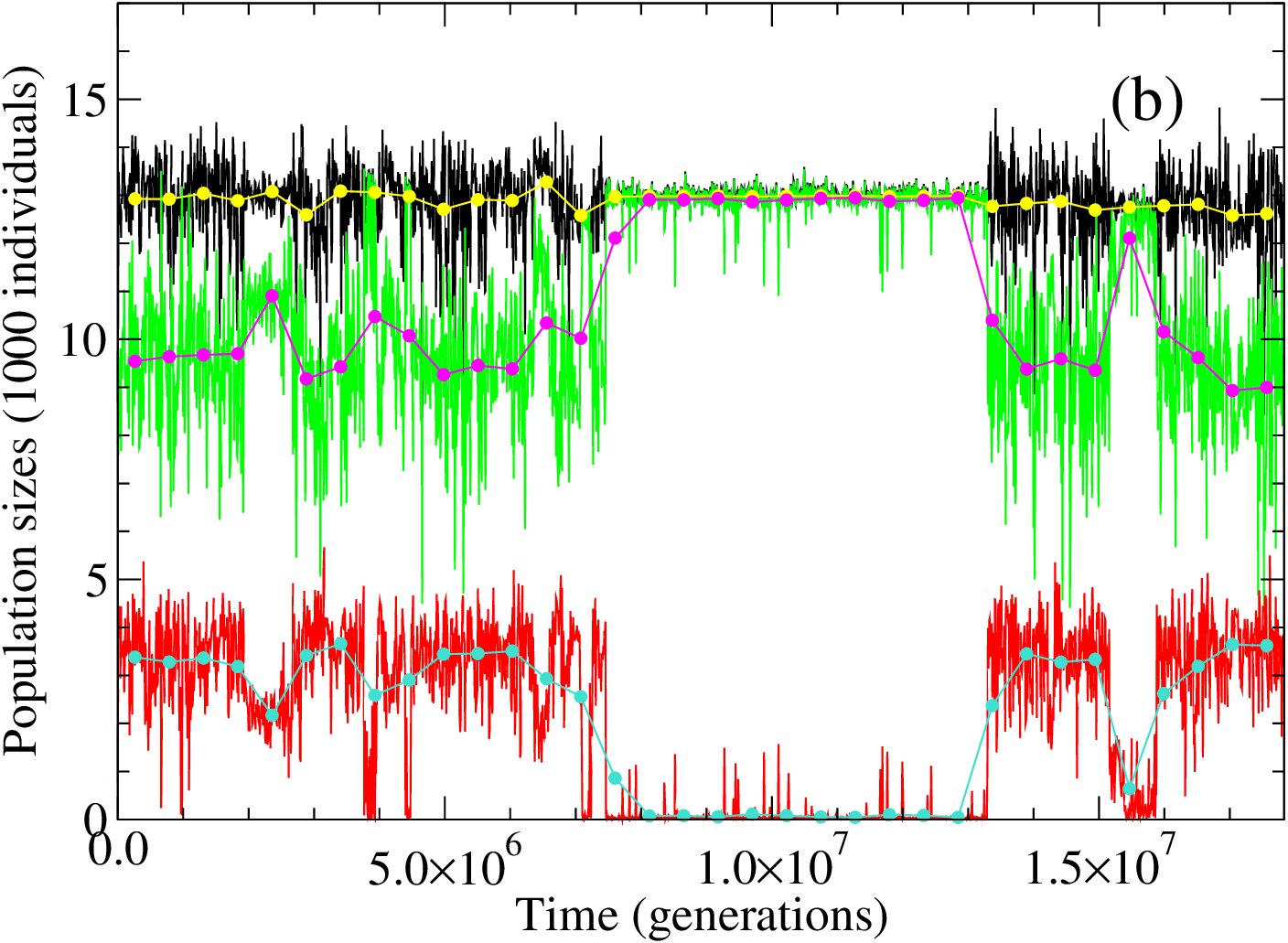}
\end{center}
\caption[]{
Time series of diversities {\bf (a)}
and population sizes {\bf (b)} for
one specific simulation run of the functional-response model without 
adaptive foraging. The
strongly fluctuating curves in the background are sampled every 
8192 generations, while the smooth curves with data points in
contrasting colors that are overlaid 
in the foreground are running averages over
524\,288 generations. Black with light gray (yellow online) 
overlay: all species.  
Light gray with dark gray overlay (green and magenta online): producers. 
Dark gray with light gray overlay (red and cyan online): consumers. 
}
\label{fig:timser}
\end{figure}
\begin{figure}[ht]
\begin{center}
\vspace*{0.4truecm}
\includegraphics[angle=0,width=.47\textwidth]{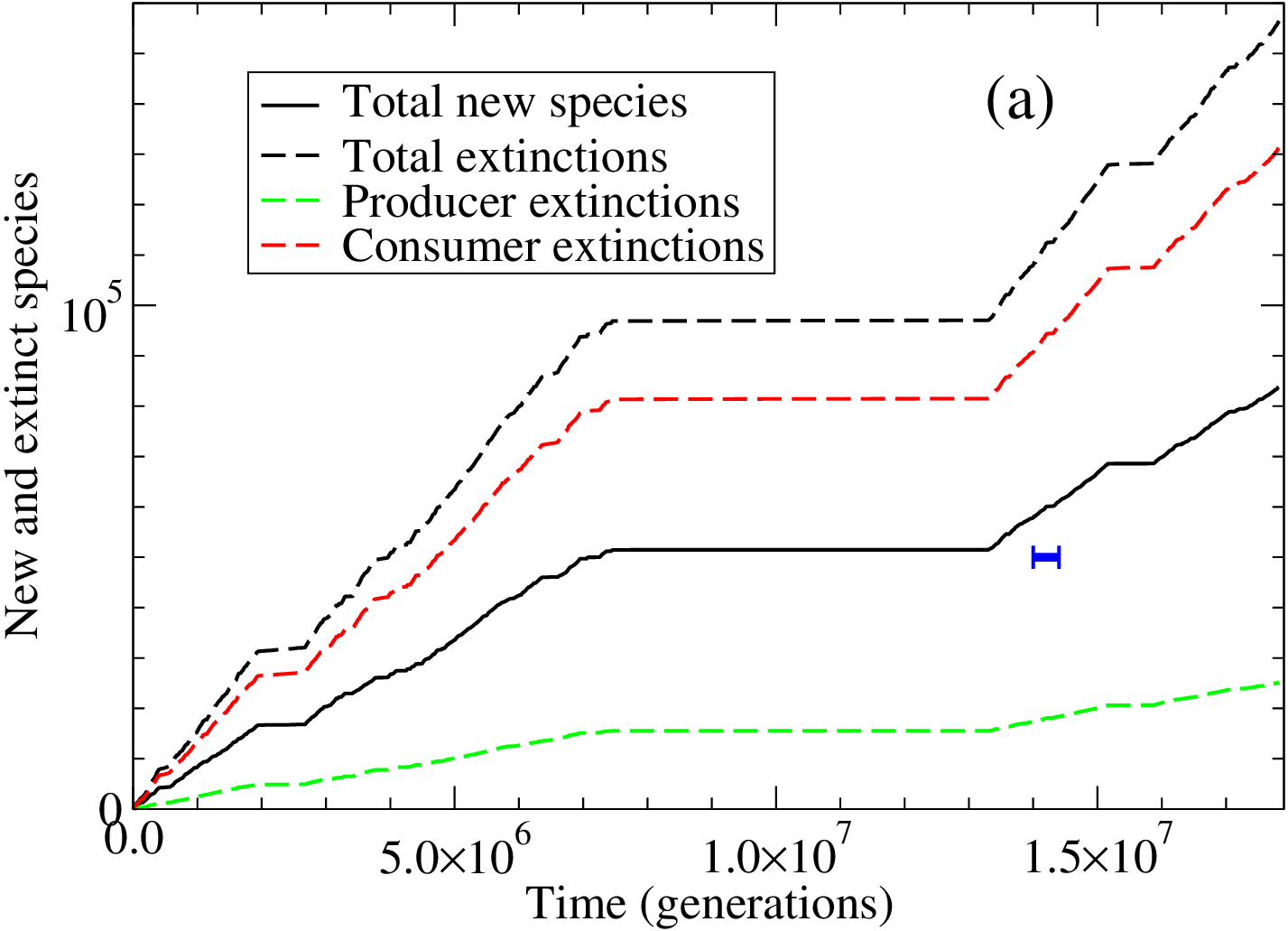}
\hspace{0.5truecm}
\includegraphics[angle=0,width=.47\textwidth]{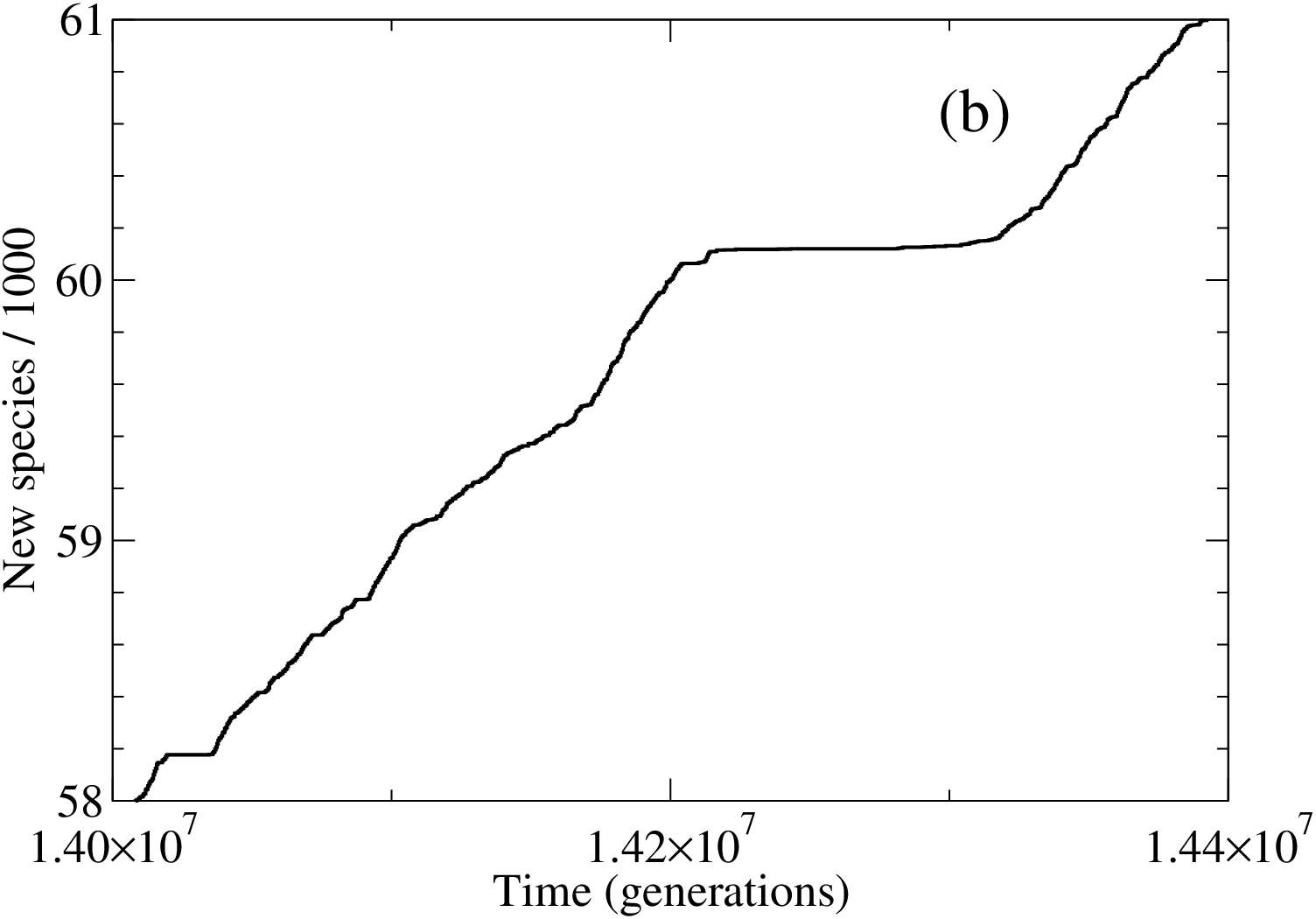}
\end{center}
\caption{
{\bf (a)}
Time series of the 
accumulation of new and extinct species in the same simulation run
depicted in Fig.~\protect\ref{fig:timser}. 
The solid, black curve shows the total number of different
species that have at least once 
attained a population size $n_I > 1000$ by time $t$. The dashed curves
count the total number of species that have gone extinct after 
attaining a maximum population greater than 1000. The black dashed
curve refers to all species, the light gray one (green online) to
producers, and the dark gray one (red online) to consumers. The
ratio of approximately 1.89 between the dashed and full black curves
indicate that major species recur on average about twice
during the evolution. This is an artifact of the finite genome
length. 
{\bf (b)}
The detailed, intermittent structure of new species 
(the solid, black curve in {\bf (a)}) over $400\,000$ generations. 
The interval is indicated by the short, horizontal bar in {\bf (a)}.  
}
\label{fig:timser2}
\end{figure}

\section{Numerical Results for the Functional-response Model}
\label{sec:Sim1}

We simulated the functional-response model defined in Sec.~\ref{sec:holl} over
$2^{24} = 16\,777\,216$ generations (plus $2^{20}$
generations ``warm-up") for the
following parameters: genome length $L=21$ 
($2^{21} = 2\,097\,152$ potential
species), external resource $R=16\,000$, fecundity $F=2$,  
mutation rate $\mu = 10^{-3}$, proportion of
producers $c_{\rm prod} =0.05$, interaction matrix $\bf M$ with
connectance $C = 0.1$ and nonzero elements with a symmetric, 
triangular distribution over $[-1,+1]$, and $\lambda = 1.0$. 
(The high value of $\lambda$ is of course biologically unrealistic, and it was 
chosen to obtain a larger population of heterotrophs for a computationally 
manageable autotroph population.)
We ran five independent runs, each starting from 100 individuals
of a single, randomly chosen producer species. 

\subsection{Time series}
\label{sec:timser}

Time series of diversities (effective numbers of species)
and population sizes 
for one realization are shown in Fig.~\ref{fig:timser}. To
filter out noise from low-population,  
unsuccessful mutations, we define the diversity as the exponential
Shannon-Wiener index \citep{KREB89}. 
This is the exponential function 
of the information-theoretical entropy of the population distributions,
$D(t) = \exp \left[S \left( \{ n_I(t) \} \right) \right]$, where
$S\left( \{ n_I(t) \} \right)
=
- \sum_{\{I | \rho_I(t) > 0 \}} \rho_I(t) \ln \rho_I(t)
$
with
$\rho_I(t) = n_I(t) / N_{\rm tot}(t)$ for the case of all species,
and analogously for the producers and consumers separately. 

The time series for both diversities and population sizes display
intermittent behavior with quiet periods of varying lengths,
separated by brief periods of high evolutionary
activity. The intermittency is highlighted by the time series for the 
accumulation of new and extinct species, shown in Fig.~\ref{fig:timser2}.
In this respect, the results are similar to those seen for the 
simplified tangled-nature models in earlier 
work \citep{RIKV05A,RIKV06,RIKV03,SEVI06}. However, 
diverse communities in this model are less stable than those
produced by the simplified tangled-nature models. 
It is possible that this instability is related to the tendency of ``triangles" 
consisting of two species competing for a common resource while one of them 
also feeds on the other one (intraguild predation) 
to collapse, that we discussed in Sec.~\ref{sec:compet}. 
The instability expresses itself in a tendency for this model 
to flip randomly between an active phase with a diversity near 
ten, and a ``garden of Eden'' phase of one or a few producer species with
a very low population size of numerous unstable consumer species, 
such as the one seen around 10 million generations in Figs.~\ref{fig:timser} 
and \ref{fig:timser2}(a).

\subsection{Power-spectral densities}
\label{sec:psd}

\begin{figure}[t]
\begin{center}
\vspace*{0.3truecm}
\includegraphics[angle=0,width=.47\textwidth]{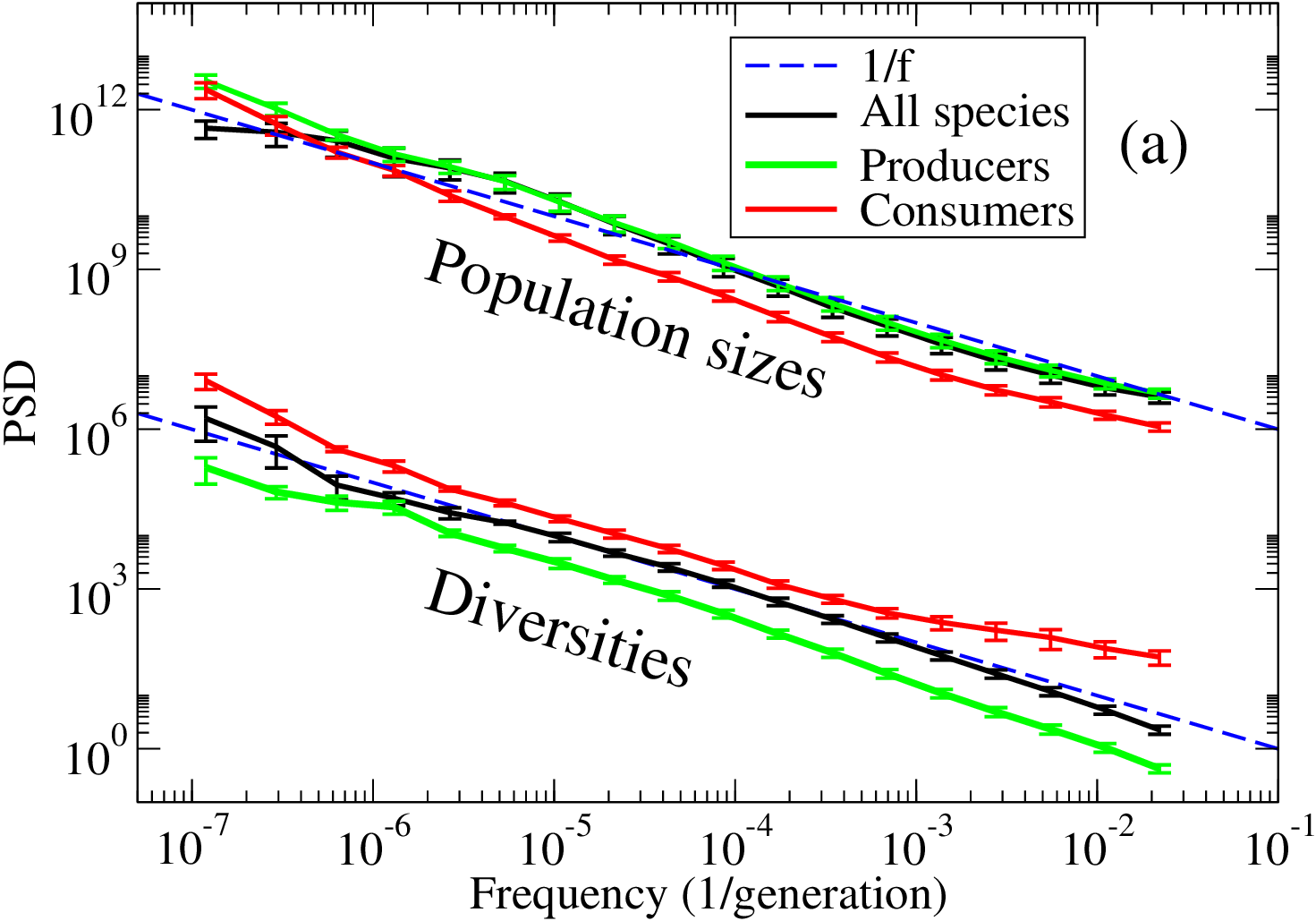}
\hspace{0.5truecm}
\includegraphics[angle=0,width=.47\textwidth]{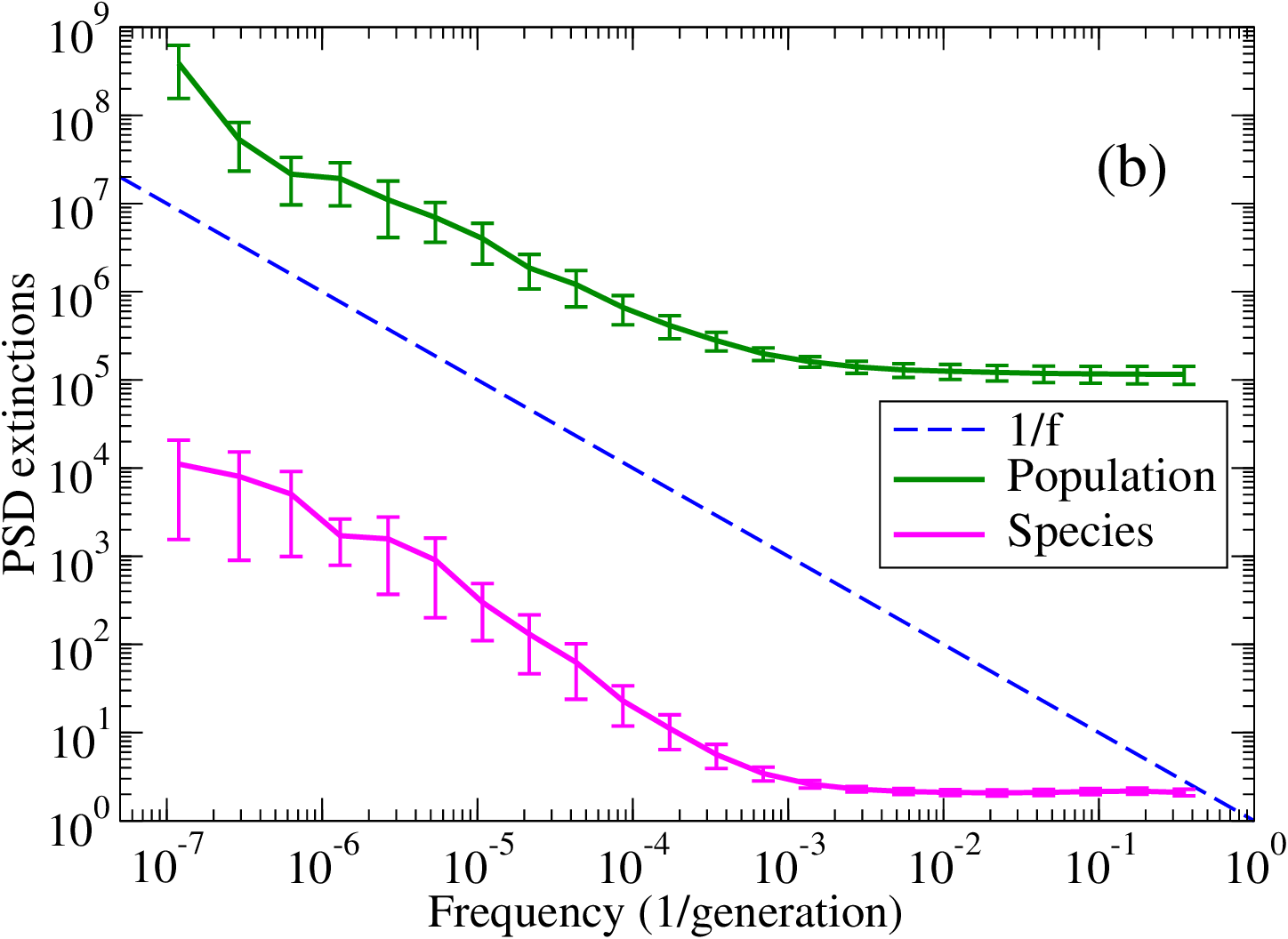}
\end{center}
\caption[]{
{\bf (a)}
PSDs for the diversities and population sizes, each recorded
separately for all species and for producers and consumers. 
The time series were sampled every 16 generations. 
{\bf (b)}
PSDs for the extinction activity, sampled every generation. 
In both parts of the figure, the results are averaged over five
independent simulation runs. 
See discussion in the text. 
}
\label{fig:psd}
\end{figure}
To obtain information about the intensity of
fluctuations in the evolving community, we calculate  
power-spectral densities, or 
PSDs.\footnote{The PSDs were calculated by obtaining the averages over octaves 
in frequency of the periodograms (squared Fourier transforms) of the time 
series, and then averaging these over the independent 
simulation runs. The error bars are standard errors, based on the spread 
between runs.}
These are
presented in Fig.~\ref{fig:psd} for the diversities and the
population sizes (Fig.~\ref{fig:psd}(a))
and the intensity of extinction events (Fig.~\ref{fig:psd}(b)). 
The former two are shown for the total population, as well as 
separately for the producers and consumers. All three are similar.
Extinction events are recorded as
the number of species that have attained a population size greater
than one, which go extinct in generation $t$ (marked as ``Species"
in the figure), while
extinction sizes are calculated by adding the maximum population sizes
attained by all species that go extinct in generation $t$ (marked
as ``Population" in the figure). 
The PSDs for all the quantities shown exhibit approximate $1/f$
behavior. For the diversities and population sizes, this power law
extends over more than five decades in time. The extinction
measures, on the other hand, have a large background of white noise
for frequencies above $10^{-3}$ generations$^{-1}$, 
probably due to the high rate of  
extinction of unsuccessful mutants. For lower frequencies, however,
the behavior is consistent with $1/f$ noise within the limited
accuracy of our results. 
We note that the apparent $1/f$ behavior in the PSD of extinction events 
in the model of \citet{NUNE99} extends over only one decade in frequency. 

\subsection{Species lifetimes and durations of quiet periods}
\label{sec:times}

\begin{figure}[t]
\begin{center}
\vspace*{0.3truecm}
\includegraphics[angle=0,width=.47\textwidth]{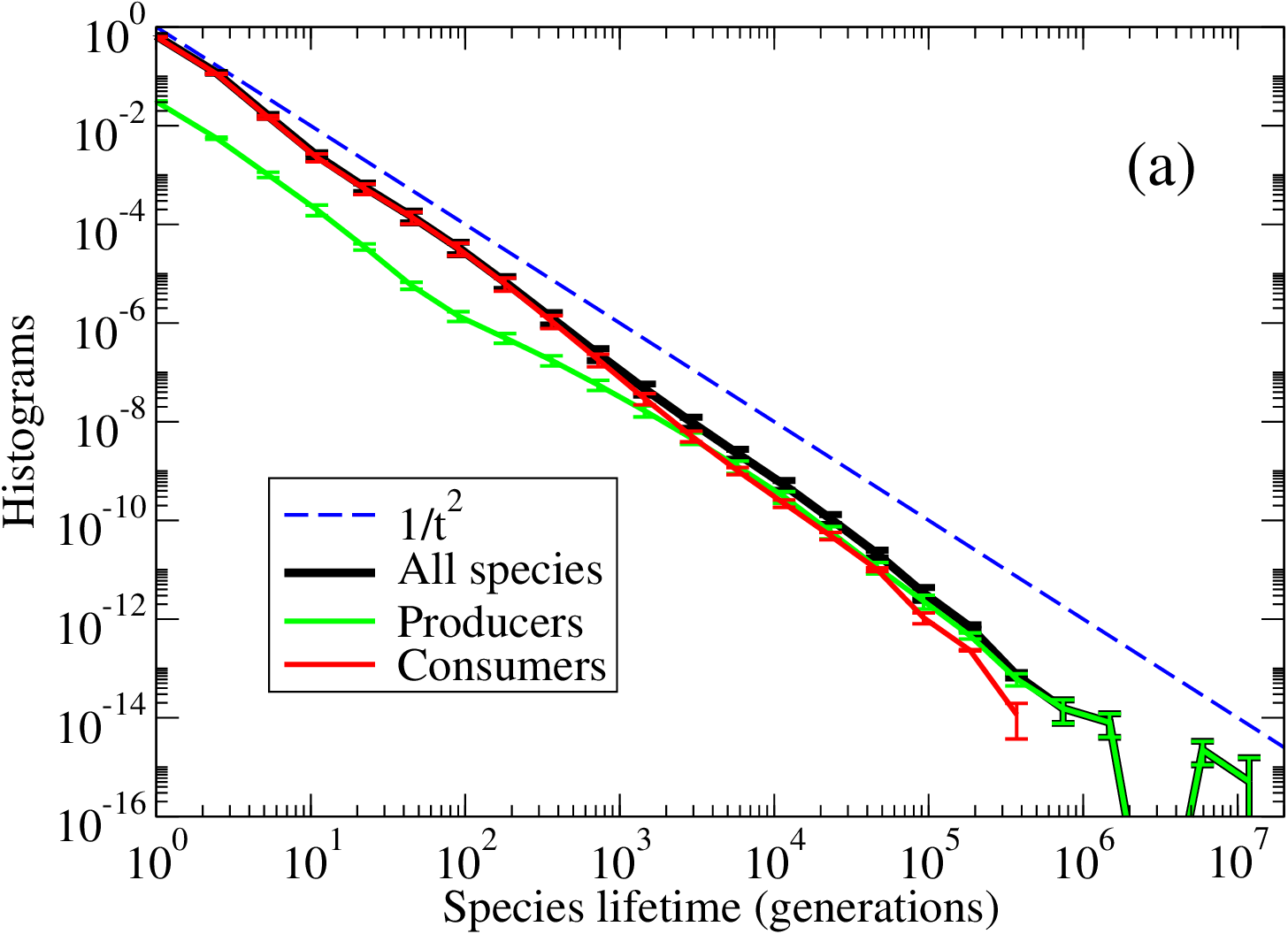}
\hspace{0.5truecm}
\includegraphics[angle=0,width=.47\textwidth]{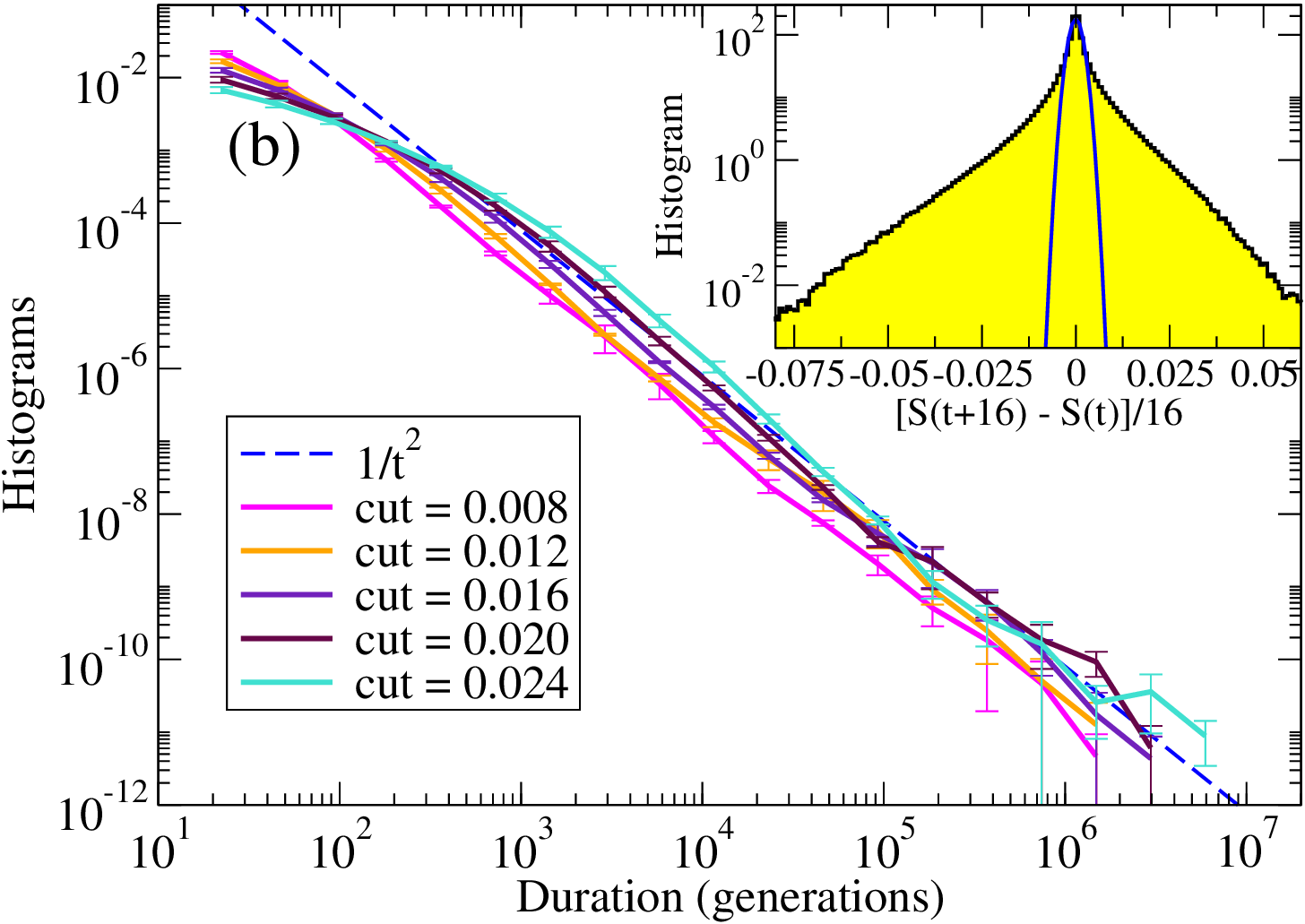}
\end{center}
\caption[]{
{\bf (a)}
Histograms of species lifetimes, shown for all species, as well as
separately for producers and consumers. 
{\bf (b)}
Histograms of the durations of evolutionarily quiet periods,
defined as the times that the logarithmic derivative of the diversity, 
$|{\rm d}S/{\rm d}t|$ (averaged over 16 generations),
falls continuously below some cutoff. The inset is a histogram of 
${\rm d}S/{\rm d}t$, showing a Gaussian center with approximately exponential
wings. The parabola in the foreground is a Gaussian fit to this
central peak. The cutoff values for the main figure, 
between 0.008 and 0.024, were chosen on the basis of this distribution. 
The data in both parts of the figure
are averaged over five independent simulation runs, and the error bars 
represent standard errors based on the differences between runs. 
}
\label{fig:time}
\end{figure}
The evolutionary dynamics can also be characterized by histograms
of characteristic time intervals, such as the time from emergence till
extinction of a species (species lifetimes) or the time intervals
during which  
some indicator of evolutionary activity remains continuously below 
a chosen cutoff (duration of evolutionarily quiet periods). 
Histograms of species lifetimes are shown in Fig.~\ref{fig:time}(a). 
As our indicator of evolutionary activity we use the magnitude of the
logarithmic derivative of the diversity, $|{\rm d}S/{\rm d}t|$, 
and histograms for the resulting durations of
quiet periods, calculated with different cutoffs, are shown in 
Fig.~\ref{fig:time}(b). Both quantities display approximate
power-law behavior with an exponent near $-2$, consistent with the
$1/f$ behavior observed in the PSDs \citep{RIKV03,PROC83}.  
It is interesting to note that the distributions for these two
quantities for this model have approximately the same exponent.
This is consistent with the previously studied, mutualistic 
model \citep{RIKV03,RIKV05A,RIKV06}, 
but not with the predator-prey model \citep{RIKV05A,RIKV06,RIKV06B}. 
We believe the linking of the power laws for the species lifetimes
and the duration of quiet periods indicate that the communities
formed by the present model are relatively fragile, so that all member
species tend to go extinct together in an avalanche-like ``mass extinction." 
In contrast, the previously studied predator-prey model
produces simple food webs that are much more
resilient against the loss of a few species, and as a result the
distribution of quiet-period durations decays with an exponent near 
$-1$ \citep{RIKV05A,RIKV06,RIKV06B}.

\section{Model with Adaptive Foraging}
\label{sec:adap}

\begin{figure}[t]
\begin{center}
\vspace*{0.4truecm}
\includegraphics[angle=0,width=.47\textwidth]{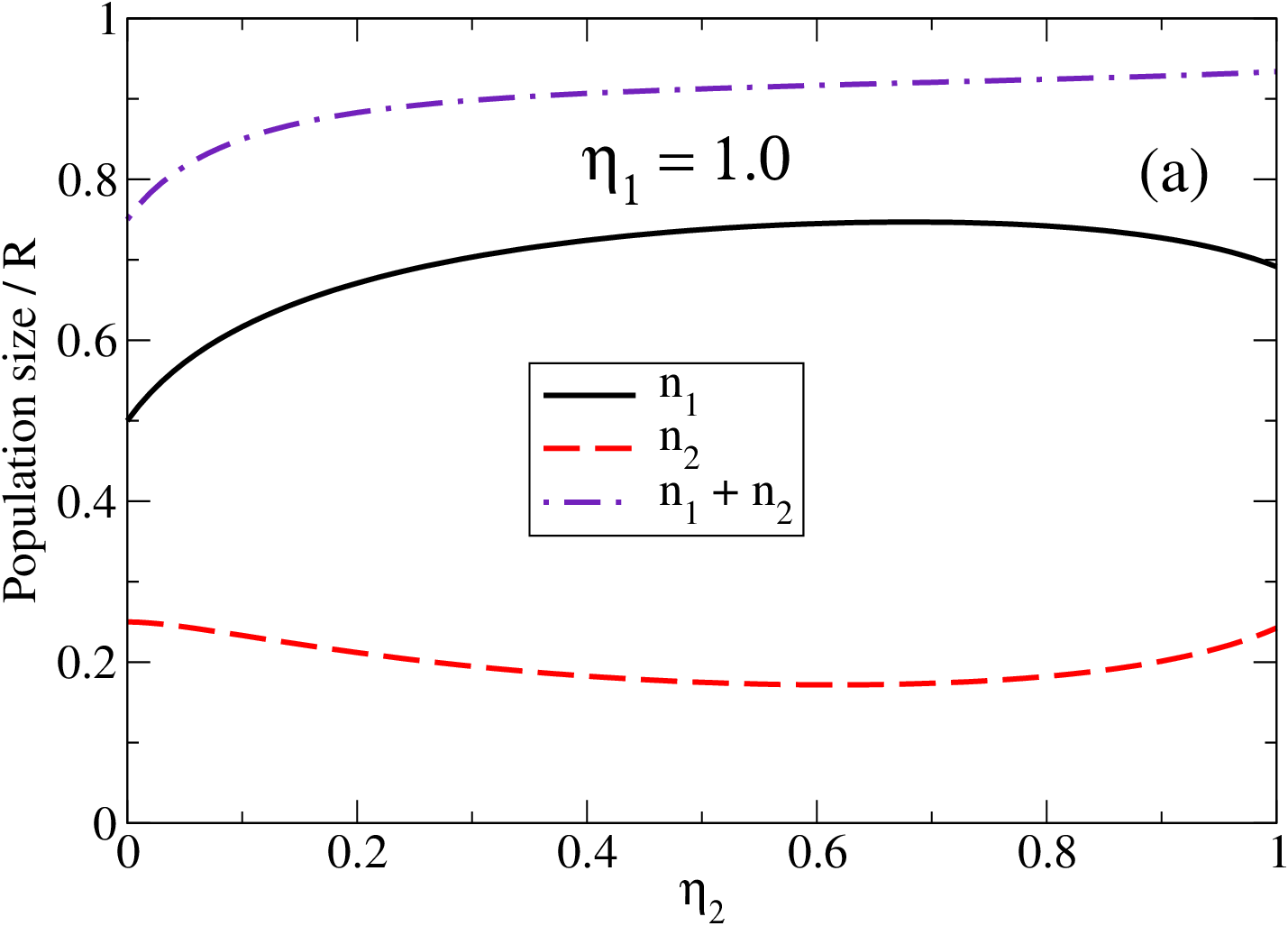}
\hspace{0.5truecm}
\includegraphics[angle=0,width=.47\textwidth]{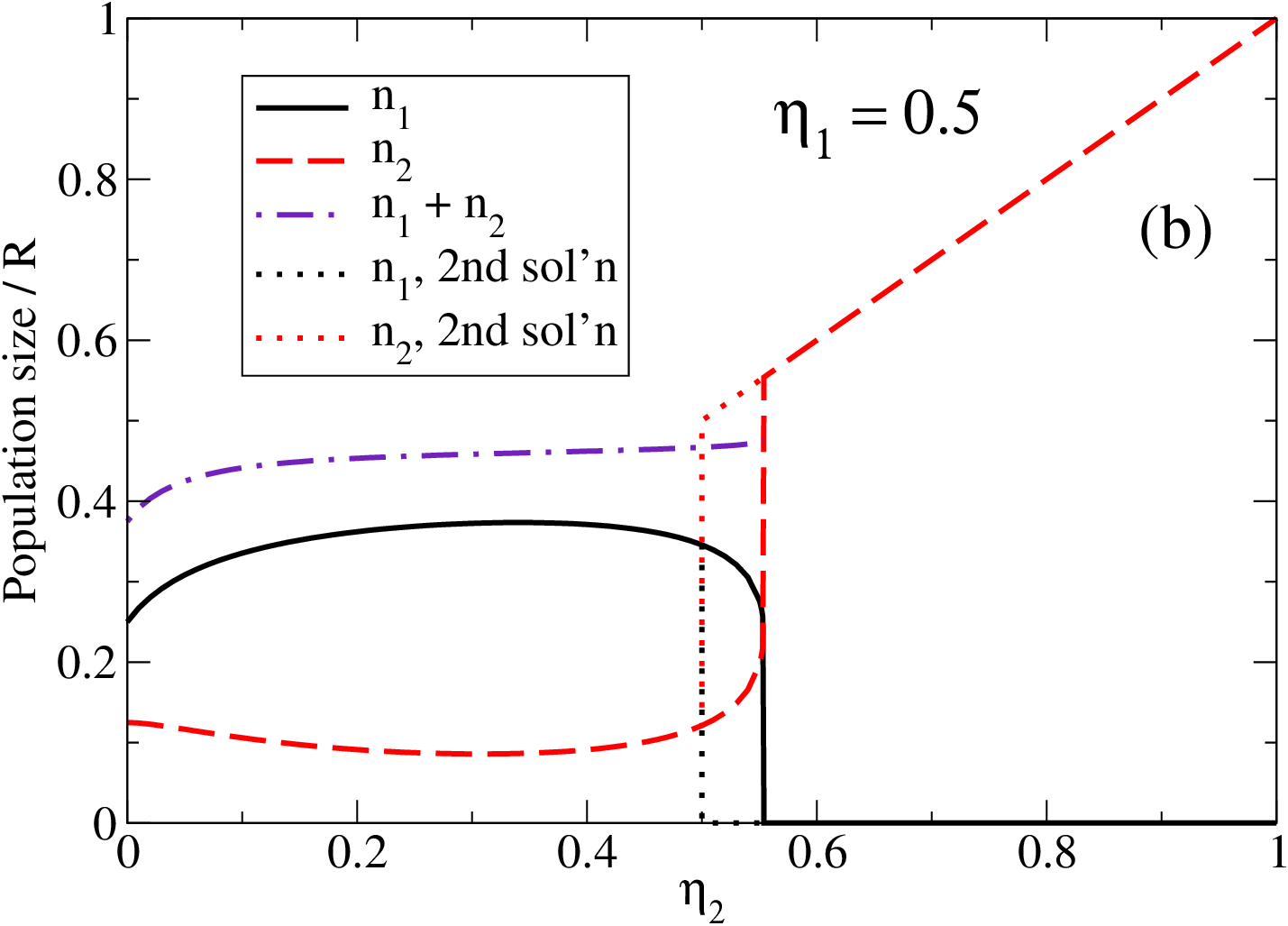}
\end{center}
\caption[]{
Population sizes for two competing producer species with intraguild 
predation in the model with adaptive foraging, shown as functions of $\eta_2$. 
{\bf (a)}
$\eta_1 = 1.0$.  
{\bf (b)} 
$\eta_1 = 0.5$.  
In both cases, $\lambda = 1$ and $M_{21} = 0.5$. 
See further discussion in the text. 
}
\label{fig:adapA}
\end{figure}
The model studied above is one in which species forage
indiscriminately over all available resources, with the output only
limited by competition. Also, there is an implication that an
individual's total foraging effort increases proportionally with
the number of species to which it is connected by a positive  
$M_{IJ}$. A more realistic picture would be that an individual's
total foraging effort is constant and can either be divided
equally, or concentrated on richer resources. The latter constitutes
adaptive foraging. 
While one can go to considerable length devising optimal foraging
strategies \citep{DROS01B,DROS04,QUIN05B,QUIN05}, we here
only use a simple scheme, in which individuals of $I$ show a
preference for prey species $J$, based on the interactions and
population sizes (uncorrected for interspecific competition).
The proportion of its foraging effort that $I$ allots to $J$ is approximated as 
\begin{equation}
p_{IJ} = \frac{M_{IJ}n_J}{\eta_I R + \sum_K^{{\rm prey}(I)} M_{IK} n_K}
\;,
\label{eq:gij}
\end{equation}
and analogously for the effort assigned to the external resource,
\begin{equation}
p_{IR} = \frac{\eta_{I} R}{\eta_I R + \sum_K^{{\rm prey}(I)} M_{IK} n_K}
\;. 
\label{eq:gir}
\end{equation}
The total foraging effort is thus normalized:
$p_{IR} + \sum_J^{{\rm prey}(I)} p_{IJ} = 1$. 
These preference factors are used to modify the reproduction 
probabilities by replacing all occurrences of $M_{IJ}$
by $M_{IJ} p_{IJ}$ and of $\eta_I$ by $\eta_I p_{IR}$ in 
Eqs.~(\ref{eq:neff}--\ref{eq:PhiIR}). 

The adaptive foraging obviously has no effect on a simple food chain 
since no species in this case has more than one choice of prey. 
The analytical results thus remain as discussed in Sec.~\ref{sec:chain}. 

For the case of two competing producers with intraguild predation 
we did not obtain 
analytical results for the fixed-point population sizes, except for the 
special cases of $\eta_2=0$, which corresponds to a two-species 
food chain, and of $M_{21}=0$, which reduces to simple competitive exclusion 
of the species with the lower $\eta_I$. 
However, numerical results for $M_{21}=0.5$, obtained by iteration of 
Eq.~(\ref{eq:MF}) with $\mu = 0$, are given in Fig.~\ref{fig:adapA}.
The results are similar for other values of the model parameters. 
The parameters in Fig.~\ref{fig:adapA}(a) are the same as in 
Fig.~\ref{fig:levels}(a), and we see that the regime of two-species coexistence 
is significantly extended by the adaptive foraging and here covers the 
full range of $\eta_2$. 
We also see that while $n_2^*$ is reduced, compared to the case without 
adaptive foraging, both $n_1^*$ and the total population, $n_1^* + n_2^*$, 
are significantly increased, indicating a more efficient overall resource 
utilization by the community. 
In Fig.~\ref{fig:adapA}(b) we reduce 
$\eta_1$ to 0.5 to explore the possibility that $\eta_2 > \eta_1$. 
We find that the coexistence solution extends up to 
$\eta_2 \approx 0.553$, where $n_1^* = n_2^*$ and the solution changes 
discontinuously to the familiar $n_2^* = \lambda (F-1) \eta_2 R$ and $n_1^*=0$. 
In fact, for $\eta_2$ between $\eta_1 = 0.5$ and 0.553, {\it both\/} 
solutions are locally stable under small perturbations. This is indicated 
by the dotted lines in the figure. Outside this range, the solutions shown 
are globally stable attractors. 
As in the case without adaptive foraging, the solution 
$n_1^*=\lambda \eta_1 (F-1)R$ and $n_2^* = 0$ is repulsive under 
perturbations to $n_2^*$. 

\begin{figure}[t]
\begin{center}
\vspace*{0.4truecm}
\includegraphics[angle=0,width=.47\textwidth]{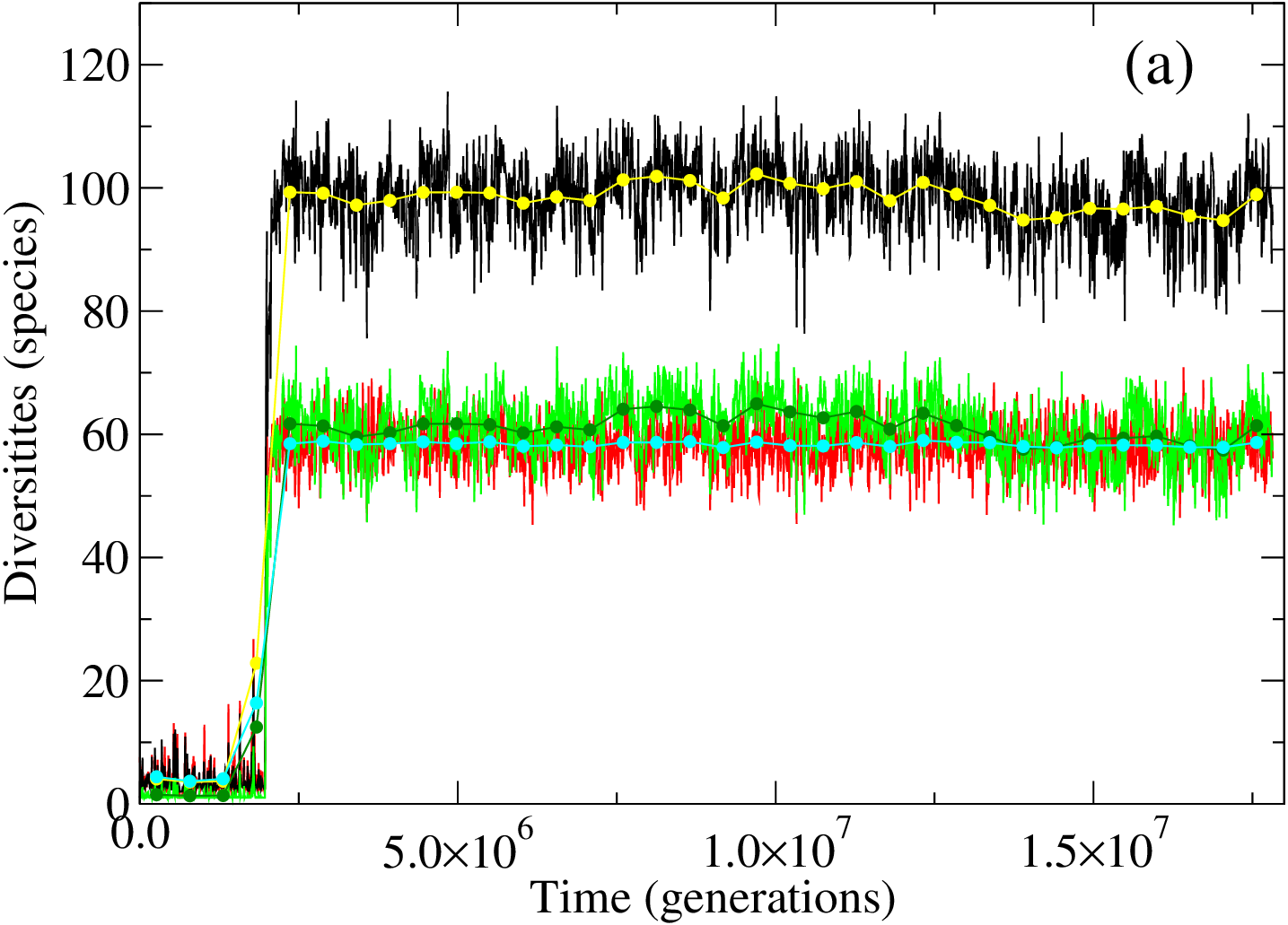}
\hspace{0.5truecm}
\includegraphics[angle=0,width=.47\textwidth]{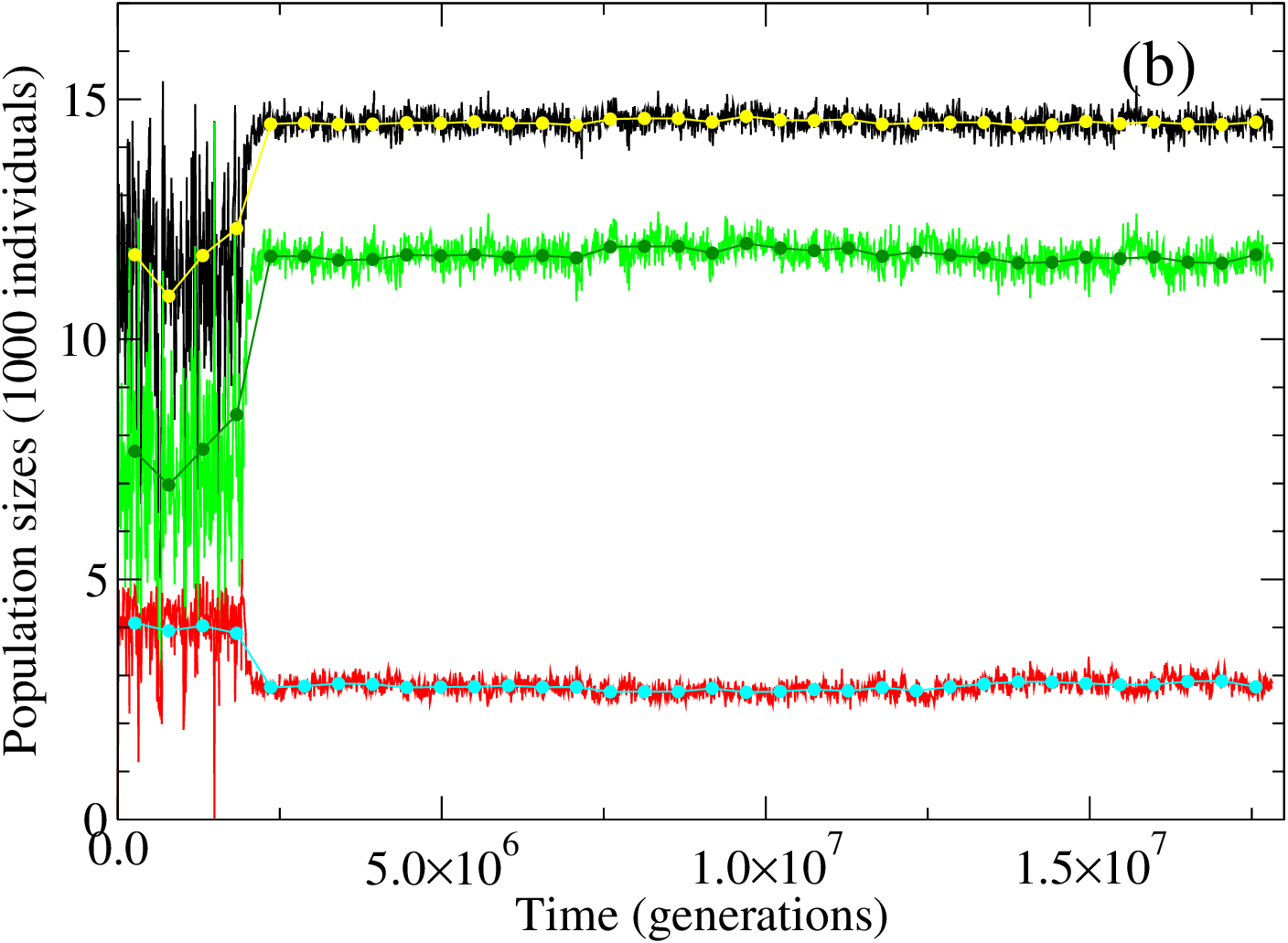}
\end{center}
\caption[]{
Time series of 
diversities {\bf (a)} and population sizes {\bf (b)} for the model
with adaptive foraging. The interpretation of the colors and lines are
the same as in Fig.~\protect\ref{fig:timser}. 
}
\label{fig:adap}
\end{figure}
\begin{figure}[ht]
\begin{center}
\vspace*{0.7truecm}
\includegraphics[angle=0,width=.47\textwidth]{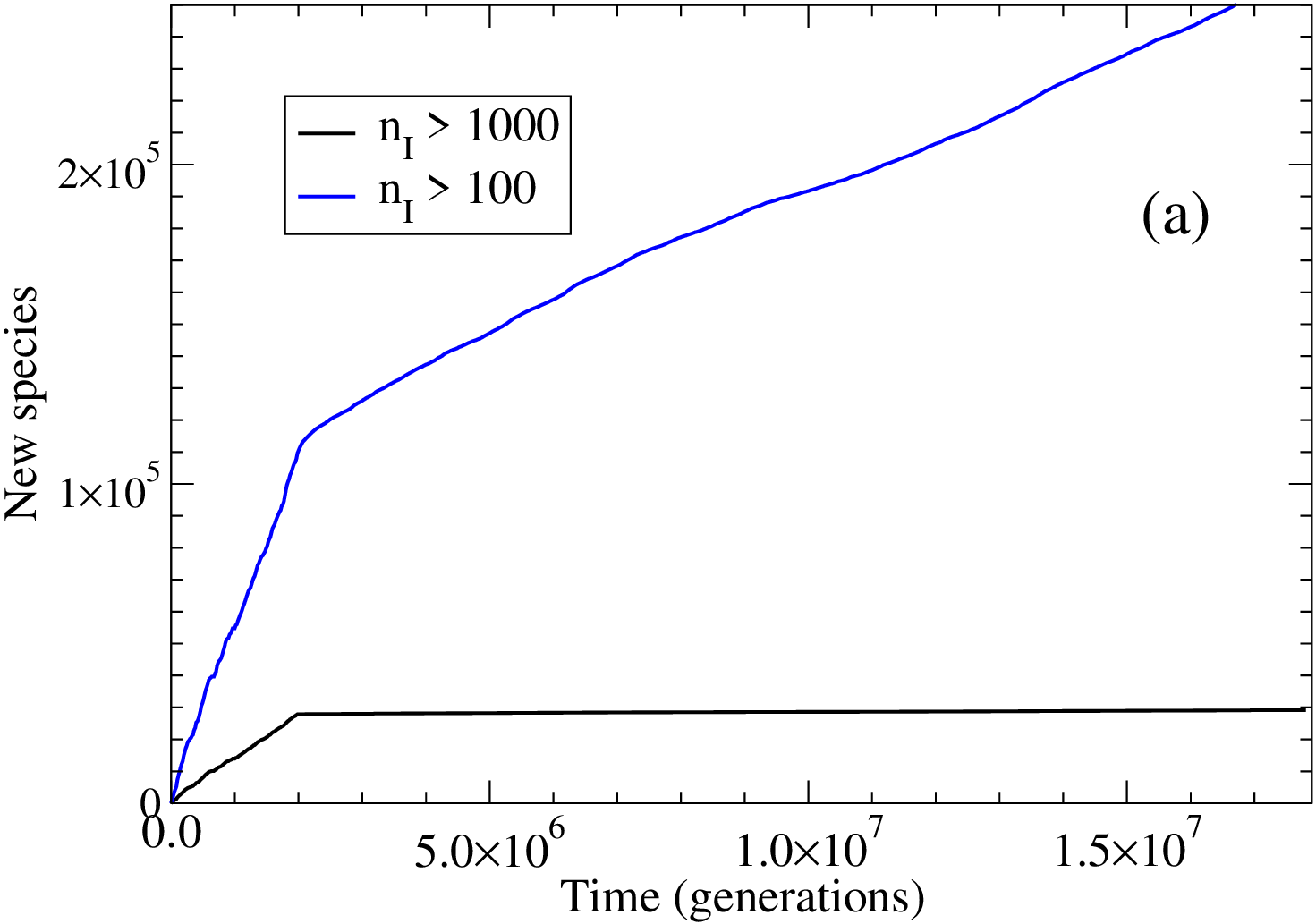}
\hspace{0.5truecm}
\includegraphics[angle=0,width=.47\textwidth]{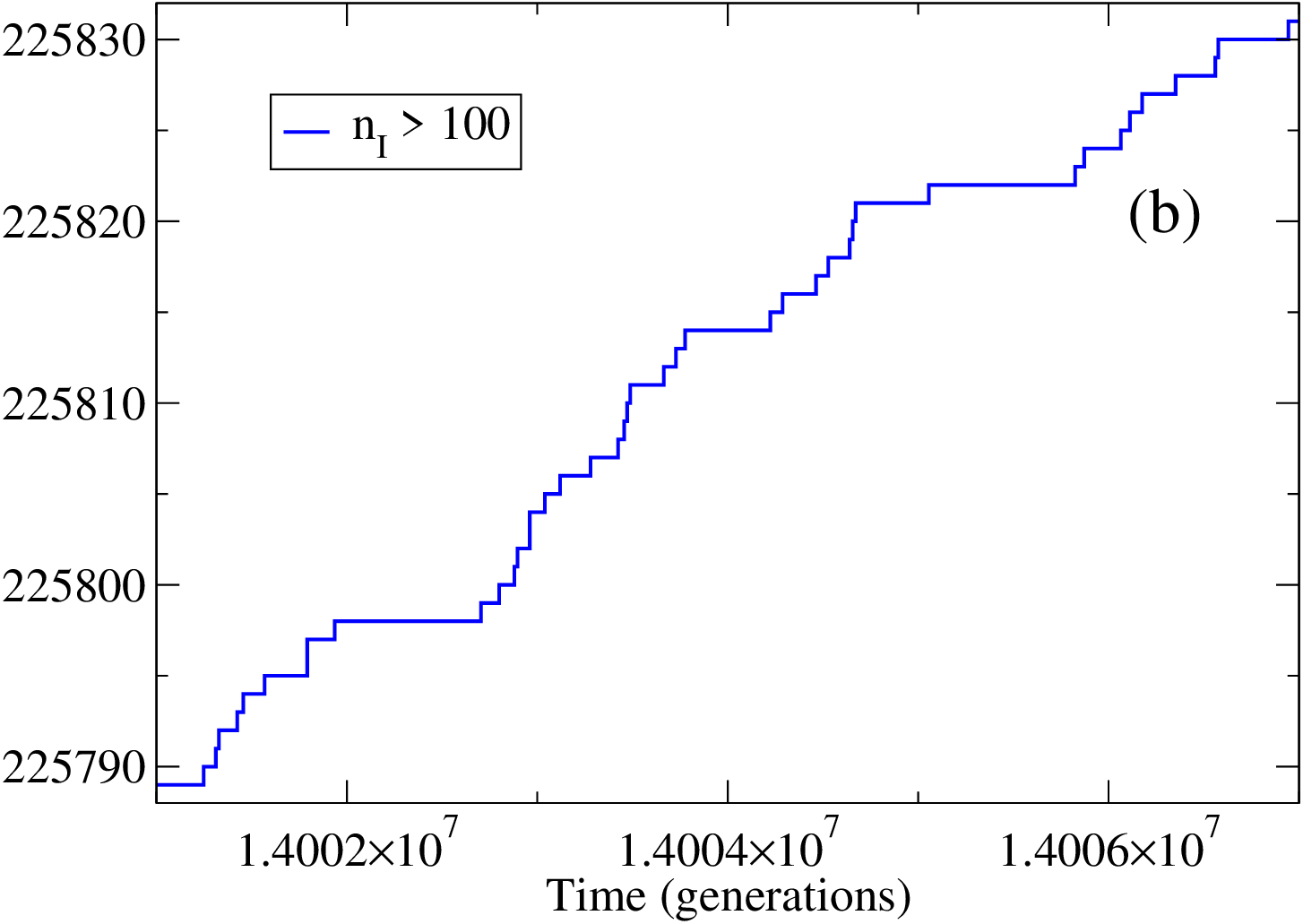}
\end{center}
\caption[]{
{\bf (a)}
Time series of 
the number of new species that have reached a population greater than
1000 (lower curve) and greater than 100 (upper curve) 
in the same simulation run depicted in Fig.~\protect\ref{fig:adap}. 
{\bf (b)}
The detailed, intermittent structure of the upper curve 
in {\bf (a)} on a very fine scale of
6000 generations.  See discussion in the text. 
}
\label{fig:adap2}
\end{figure}

The results of implementing the adaptive foraging strategy in long-time 
simulations of evolving multispecies communities are quite striking.
The system now has a metastable low-diversity phase 
similar to the active phase of the non-adaptive model, from
which it switches at a random time to a stable
high-diversity phase with much smaller fluctuations. 
As seen in Fig.~\ref{fig:adap}, the switchover
is quite abrupt, and Fig.~\ref{fig:adap2} shows that it is accompanied by
a sudden reduction in the rate of emergence of new species.
The existence of a stable high-diversity phase 
with increased producer and total populations and reduced consumer population 
in the case of adaptive foraging is consistent with the extension of the 
stability of two-species coexistence discussed above. The increased total 
population is consistent with the improved resource utilization, and the sudden 
nature of the transition from low to 
high diversity is consistent with the discontinuity and bistability, 
all observed in the adaptive two-species case.
As adaptive foraging implies an effective reduction of omnivory, our results 
present a scenario in which reduced omnivory leads to increased 
community stability \citep{NAMB08}.  

\begin{figure}[t]
\begin{center}
\vspace*{0.2truecm}
\includegraphics[angle=0,width=.47\textwidth]{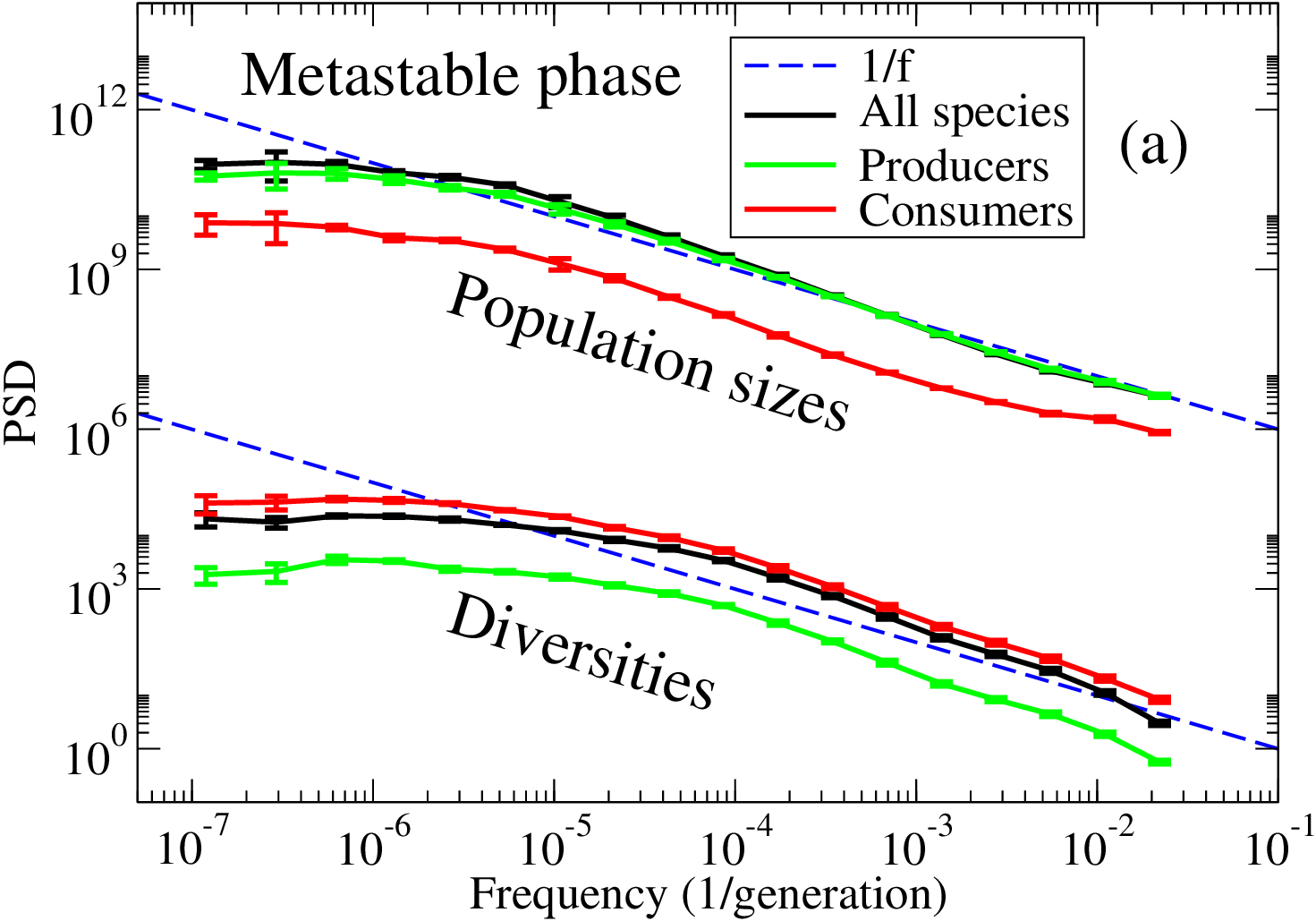}
\hspace{0.5truecm}
\includegraphics[angle=0,width=.47\textwidth]{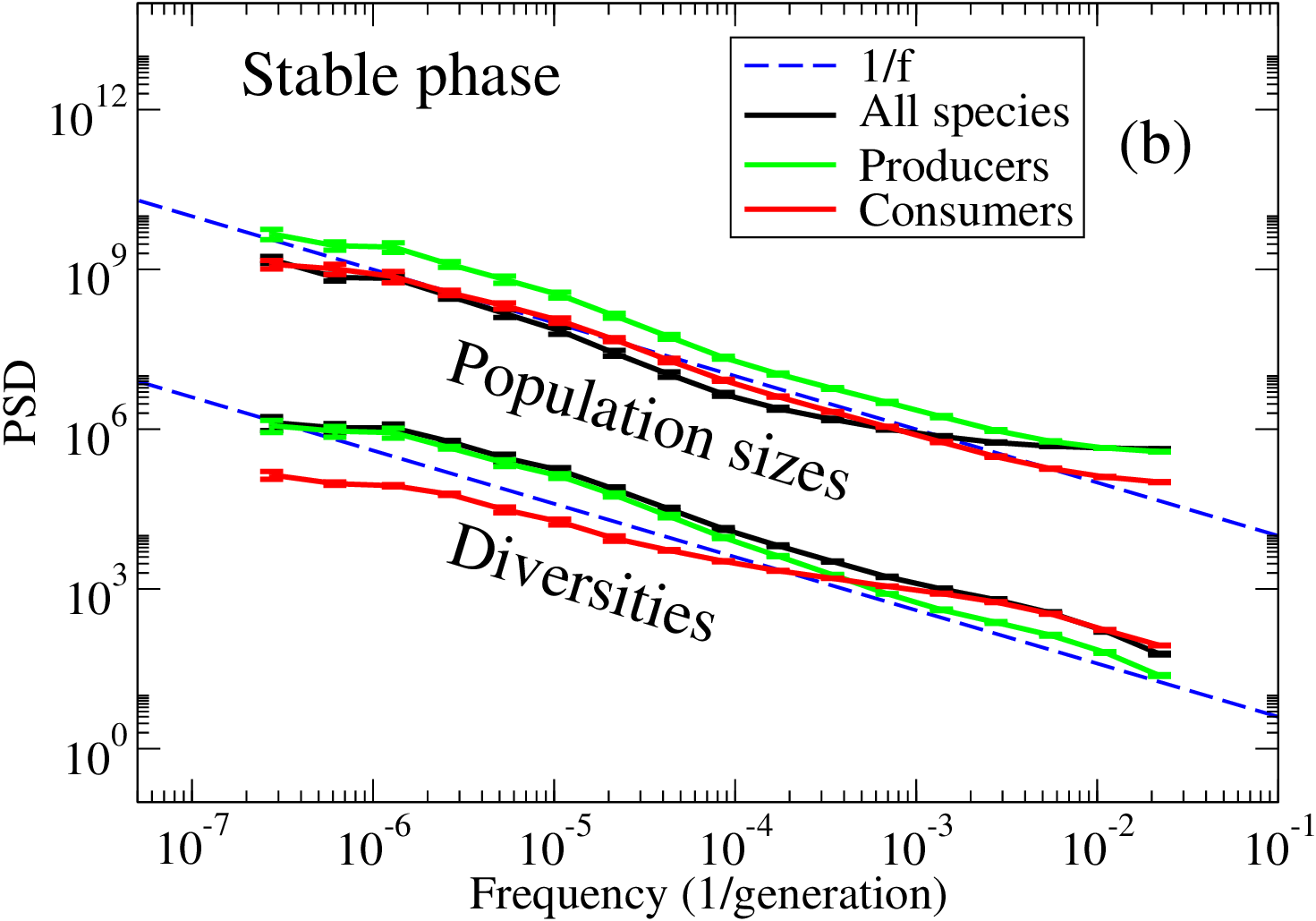}
\end{center}
\caption[]{
PSDs for the diversities and population sizes in the metastable phase
(averaged over three independent runs)
{\bf (a)}, and the stable phase (averaged over five independent runs)
{\bf (b)} for the model with adaptive foraging. Both show
approximate $1/f$ noise for frequencies above about $10^{-5}$. 
However, in the metastable phase 
the PSDs appear to approach constant levels for the lowest frequencies, 
indicating the absence of long-time correlations. 
}
\label{fig:adapPSD}
\end{figure}

As seen in Fig.~\ref{fig:adapPSD}, the PSDs for both the diversities and
population sizes in both phases show approximate $1/f$ noise for
frequencies above $10^{-5}$ generations$^{-1}$. For lower
frequencies, the metastable phase shows no discernible frequency
dependence, while for the stable phase, the frequency dependence continues 
at least another decade. It thus appears that long-time correlations are
not seen beyond about $10^5$ generations in the metastable phase. 

In fact, the system can also escape from the
low-diversity phase to total extinction, which is an absorbing state,
and in some of our simulation runs we avoided this by limiting
$|M_{IJ}|$ to less than 0.9. This restriction does not seem to have any
effect on the dynamics in the high-diversity phase.

\section{Conclusions}
\label{sec:conc}

In this paper we have extended our study of the long-time dynamics of 
a class of individual-based models with stochastic population dynamics, 
nonoverlapping generations, and random mutations during reproduction. 
Previous studies concentrated on simplified versions of the tangled-nature 
model \citep{CHRI02,COLL03,HALL02}, 
sharing with that model a reproduction probability 
defined by Eqs.~(\ref{eq:PI}) and (\ref{eq:Delta}) 
\citep{RIKV05A,RIKV06,RIKV06B,RIKV03,SEVI06,ZIA04}. 
While these models in the absence of mutations allow for exact, analytical 
solutions for the fixed-point populations of any given community of species, 
this convenient mathematical property is due to two somewhat unrealistic 
features. 
(i) The lumping together of resources and predation as respectively positive 
and negative contributions to the quantity $\Delta_I$ in 
Eq.~(\ref{eq:Delta}). (ii) The normalization of the population sizes $n_J$ 
in $\Delta_I$ by the total population size $N_{\rm tot}$, 
which can be seen as an indiscriminate, universal competition effect. 

Here, we therefore introduce population dynamics based on a 
functional response for predators versus their prey and for autotrophs 
versus the external resource. Interspecific competition is introduced through 
the competition-adjusted resources defined in 
Eq.~(\ref{eq:SI}), and intraspecific competition is 
accounted for by the ratio-dependent functional response due to 
\citet{GETZ84}, Eq.~(\ref{eq:PhiIJ}). The probability for a live  
individual to give birth is given by Eq.~(\ref{eq:BI}), and 
the probability that an individual 
avoids being eaten before it can reproduce is given by Eq.~(\ref{eq:AI}). 
Their product, Eq.~(\ref{eq:PI2}), is the total reproduction probability 
$P_I$. 

We considered two versions of this model, one without adaptive foraging, 
and one with it. While a complete analytic solution of the 
fixed-point communities is not feasible for either model, 
we obtained solutions for 
a simple food chain of predators supplied by a single producer 
species. For the model without adaptive foraging we also obtained an 
analytic solution for the coexistence of two producer species, one of which 
also acts as a predator toward the other (intraguild predation). 
A corresponding analytic solution was not 
found for the model with adaptive foraging. However, numerical solutions of 
Eq.~(\ref{eq:MF}) with $\mu = 0$ 
showed a significantly expanded parameter range for coexistence, including 
a regime where both coexistence and competitive exclusion are locally 
attractive fixed-point solutions. 
The coexistence solution also exhibited a decrease in the predator 
population, which was more than compensated by an increase in the prey 
population, leading to a significant increase in the total population size; 
in other words to a more efficient resource exploitation by the overall 
community. 

Long-time kinetic Monte Carlo simulations of the two models produced time 
series of diversities and population sizes that exhibit approximate 
$1/f$ noise in a wide frequency range. However, there are significant 
differences between the models. 

Without adaptive foraging, the community 
flips randomly back and forth between an evolutionarily 
active phase with diversities 
around ten, and another phase with a very low evolutionary turnover, 
a small number of coexisting producer species, and a very 
small population of unsuccessful consumers. 

With adaptive foraging, the model displays a metastable phase which resembles 
the active phase in the previous case. After a random amount of time, this 
phase suddenly gives way, either to total extinction, or to a new, stable phase 
with an order of magnitude larger diversities and somewhat higher total 
population size. The relative fluctuations in this phase are much smaller 
than in the other phases in either model. We find it reasonable to believe
that the increased diversity and population size in the stable phase of the 
model with adaptive foraging are related to the increased tendency toward 
species coexistence in this model, observed in Fig.~\ref{fig:adapA}. 
This stabilization of the communities is consistent with 
observations for the web-world model with adaptive foraging 
\citep{DROS01B,DROS04,QUIN05B,QUIN05}. 
The dynamics and community structures of the stable phase in the model with 
adaptive foraging 
are studied in detail in a forthcoming paper \citep{RIKV09}. 

In summary, the models studied here show that some aspects of the long-time 
dynamics, such as the $1/f$ noise in diversities, population sizes, and 
extinction sizes, are quite robust. However, the community structures and 
their stability or lack thereof show significant differences, both from 
previously studied 
tangled-nature type models, and depending on whether or not adaptive foraging 
is implemented. A comprehensive understanding of the 
universal and nonuniversal fluctuation properties 
of large-scale coevolution and their relation to avalanches of  
species extinctions is likely to demand a combination of 
implementations of more realistic population dynamics and mutation mechanisms, 
and further study of minimalistic models.

\section*{Acknowledgments}

Supported in part by U.S.\ National Science Foundation Grant Nos.\ 
DMR-0444051 and DMR-0802288.




\end{document}